\documentclass[superscriptaddress,aps,prd,twocolumn,preprintnumbers,nofootinbib,reprint,longbibliography]{revtex4-2}
\usepackage[utf8]{inputenc}
\usepackage{uniinput}
\usepackage{amsmath,mathtools,mathrsfs,hyperref}
\hypersetup{
	colorlinks=true,
	allcolors=blue
}

\usepackage{microtype}
\usepackage[plain]{fancyref}
\usepackage{cleveref}
\usepackage{enumitem, subfigure}
\usepackage{xcolor, transparent}
\usepackage{fourier-orns}

\usepackage[commentmarkup=uwave]{changes}
\definechangesauthor[name=SG, color=blue]{SG}

\newcommand{\phiin}{Φ_{\textrm{in}}}
\newcommand{\phishell}{Φ_{\textrm{shell}}}
\newcommand{\phiout}{Φ_{\textrm{out}}}
\renewcommand{\d}{\mathrm{d}}
\renewcommand{\hat}{\widehat}

\begin{document}

\title{
Exploring Black Hole Mimickers:\\Electromagnetic and Gravitational Signatures of AdS Black Shells}
\author{Suvendu Giri}
%\email{suvendu.giri@princeton.edu}
\affiliation{Department of Physics, Princeton University, Princeton, New Jersey 08544, USA}
\affiliation{Princeton Gravity Initiative, Princeton University, Princeton, NJ 08544, USA}
\affiliation{Institutionen för fysik och astronomi,
		Uppsala Universitet, Box 803, SE-751 08 Uppsala, Sweden}
\author{Ulf Danielsson}
%\email{ulf.danielsson@physics.uu.se}
\affiliation{Institutionen för fysik och astronomi,
		Uppsala Universitet, Box 803, SE-751 08 Uppsala, Sweden}
\author{Luis Lehner}
%\email{llehner@perimeterinstitute.ca}
\affiliation{Perimeter Institute for Theoretical Physics, 31 Caroline St., Waterloo, ON, N2L 2Y5, Canada}
\author{Frans Pretorius}
%\email{fpretori@princeton.edu}
\affiliation{Department of Physics, Princeton University, Princeton, New Jersey 08544, USA}
\affiliation{Princeton Gravity Initiative, Princeton University, Princeton, NJ 08544, USA}

\begin{abstract}
\noindent
We study electromagnetic and gravitational properties of AdS black shells (also referred to as AdS black bubbles)---a class of quantum gravity motivated black hole mimickers, that in the classical limit are described as ultra compact shells of matter. We find that their electromagnetic properties are remarkably similar to black holes. We then discuss the extent to which these objects are distinguishable from black holes, both for intrinsic interest within the black shell model, and as a guide for similar efforts in other sub-classes of exotic compact objects (ECOs). We study photon rings and lensing band characteristics, relevant for very large baseline inteferometry (VLBI) observations, as well as gravitational wave observables---quasinormal modes in the eikonal limit and the static tidal Love number for non-spinning shells---relevant for ongoing and upcoming gravitational wave observations.
\begin{center} 
    \rule{10pt}{0.2pt} {\fontencoding{U}\fontfamily{futs}\selectfont\aldine\relax}
    \rule{10pt}{0.2pt} 
\end{center}
\end{abstract}

\preprint{UUITP-12/24}

\maketitle
\allowdisplaybreaks

\tableofcontents

\section*{Introduction and overview}
The ability to scrutinize black holes has recently kicked into high gear. The detection of 
gravitational waves has opened a window to directly examine dynamical, strong field gravity, in particular from systems describing stellar mass black hole mergers~\cite{PhysRevLett.116.061102,PhysRevX.13.041039}. 
Electromagnetic signals detected through very large baseline interferometry are providing a glimpse into horizon-scale emission regions of supermassive black holes~\cite{EventHorizonTelescope:2019dse}. These exquisite detection
efforts, which will be followed by increasingly more sensitive campaigns, herald unprecedented
opportunities to test fundamental ideas about the nature of black holes, and potential deviations from
their simple structure predicted by (classical) General Relativity.

Multiple lines of arguments rooted in quantum theory call into question the classical picture.  Indeed, the
so-called black hole information paradox, and the connection of black hole entropy to its area, have motivated diverse models attempting to interpret and reconcile such
puzzling aspects of black holes.  These, and connected efforts to address the presence of singularities inside black holes, have motivated ``extensions'' to black holes, altering their structure in the vicinity of the classical horizon and its interior. Proposals for objects like fuzzballs \cite{Lunin:2002qf,Giusto:2004id,Mathur:2005zp}, boson stars \cite{Liebling:2012fv}, gravastars \cite{Mazur:2001fv,Visser:2003ge}, AdS black shells, and other alternatives have been presented (see \cite{Cardoso:2019rvt} and references therein), collectively referred to as exotic compact objects (ECOs)  (not all 
necessarily motivated by quantum gravity considerations).

Here, we focus on a subclass of ECOs, ``AdS black bubbles'' which, for
their characteristics described in this work, will be referred to as {\em
shells}. Such an object is modeled as an
ultra-compact thin-shell: a 2-sphere surface layer of matter close to the would-be horizon of the analogous black hole, but still at a macroscopic distance outside, that separates a nonsingular interior spacetime from the exterior, asymptotically flat spacetime \cite{Danielsson:2017riq, Danielsson:2017pvl, Danielsson:2021ykm, Danielsson:2021ruf, Danielsson:2023onu}.
In previous works we have focused on understanding the dynamics of black shells in spherical symmetry, metric properties of slowly rotating shells including its quadrupole moment, and discussed prospects and challenges to be addressed for testing the black shell paradigm \cite{Danielsson:2021ykm,Danielsson:2023onu}.
In this paper, short of considering binary black shell systems, we study a set of observable consequences tied to black shells in both electromagnetic and gravitational observables. 
We discuss the extent to which such objects are distinguishable from black holes, both for intrinsic interest within the black shell model, and also as a guide for similar efforts in other subclasses of ECOs.
\footnote{There is increasing interest in understanding observational properties of ECOs, with the intention of distinguishing them from black holes see e.g. \cite{Younsi:2016azx,Tamm:2023wvn,Rosa:2024bqv}. 
However, most efforts focus on electromagnetic observables in stationary regimes. This is because dynamics of most models of ECOs (with the exception of boson stars, see e.g. \cite{Liebling:2012fv,Olivares:2018abq,Siemonsen:2024snb}) are not sufficiently developed to be able to study their merger in binaries.}

In \cref{sec:review}, we begin with an overview highlighting key aspects of black shells, which will serve as the concrete model that we will study in the rest of this paper.

In \cref{sec:em_properties,sec:generator}, we focus on the electromagnetic properties of black shells. By assigning sufficiently large values to parameters determining the permittivity, permeability and conductivity of the shell, we find that the black shells satisfy a ``no-hair'' property, akin to the electromagnetic no-hair properties of Kerr-Newman black holes. We also find that they are fully absorbing, and that currents flowing within the thin shells have the same properties as the effective currents on black hole horizons. We argue that black shells can serve as electric generators producing jets in a similar way as black holes.

In \cref{sec:em_observables}, we study observational signatures of black shells in the electromagnetic spectrum, relevant for very large baseline interferometry (VLBI) observations made by the event horizon telescope (EHT). Since black shells are macroscopically bigger than black holes, this changes the size and shape of the inner shadow. Owing to the difference in quadrupole moment of a rotating black shell as compared to Kerr, the critical curve (photon ring) has a shape distinct from that of a Kerr black hole with identical mass and spin. Both of these are potentially observable effects that could distinguish a black shell from a black hole.

In \cref{sec:gw_observables}, we study observables that are relevant for gravitational wave observations of binary mergers by the LIGO-Virgo-Kagra (LVK) collaboration, as well as upcoming ground-based and space-based missions. We begin, in \cref{sec:light_ring}, by examining the characteristic frequencies associated with the light ring---the orbital frequency and the instability exponent---that can be used to estimate the quasi-normal mode frequencies in the large harmonic number $\ell$ limit. These frequencies are different from those of a Kerr black hole because of a difference in multipole moments. Unsurprisingly, the difference is small, owing to only percent level differences in the multipole moments. We then compute the tidal Love number of stationary black shells in \cref{sec:tidal}, where we find it to be small and positive. A non-zero Love number would be observable in the inspiral phase of a gravitational wave signal prior to merger, imprinting itself on the signal as an overall phase difference compared to black hole waveforms (for which the Love number vanishes).

Finally, in \cref{sec:outlook} we comment on what to expect in the gravitational wave signal from a black hole binary merger event, a discussion on other pertinent aspects of the black shell model which generalizes to ECOs, and conclude with an outlook on future directions.

\section{Black shells: an overview}\label{sec:review}

Introduced in \cite{Danielsson:2017riq}, a black shell is a bubble of four-dimensional anti-de Sitter (AdS) spacetime enclosed by a 2+1 dimensional thin material shell, outside of which there is an asymptotically flat Minkowski vacuum. This model is inspired by string theory, giving an alternative endpoint to gravitational collapse than a black hole. The construction assumes that our four-dimensional Minkowski spacetime is metastable and may transition to an AdS vacuum through a first-order phase transition, via quantum tunneling. Such AdS vacua are ubiquitous in string theory, and the decay proceeds through the nucleation of a spherical brane bubble in four dimensions. This tunneling process is highly suppressed, making the four-dimensional Minkowski vacuum extremely long lived. However, during gravitational collapse, a new possibility emerges. Should a bubble form atop a collapsing shell of matter, the matter transforms into a gas of open strings on the nucleating brane. Assuming that this gas thermalizes at the Unruh temperature, it acquires an entropy comparable to that of the corresponding black hole, significantly increasing the phase space for nucleation and rendering its creation inevitable.

It was argued in \cite{Danielsson:2017riq} that the natural radius for converting infalling matter into radiation is the \emph{Buchdahl radius} ($=9M/4$), making it the natural radius for the formation of a black shell as suggested by the entropy argument above. To ensure stability of the shell at this radius, it is necessary to allow a transfer of energy between the matter components. This was studied in detail in \cite{Danielsson:2021ykm}, where dynamical, non-linear stability in response to radial perturbations and accretion was shown to be possible for a range of parameters describing the internal flux between the matter components of the shell.

The construction was extended in \cite{Danielsson:2017pvl, Danielsson:2021ruf,Danielsson:2023onu} to include rotating black shells with a moderately large spin (up to $a^6$ perturbatively). The external spacetime of these rotating shells shows deviations from the Kerr metric, including multipole moments that vary from those of Kerr at a percent level. Additionally, the gas on top of the rotating shell exhibits significant viscous properties, such as non-zero heat conductivity and shear viscosity. Despite its stringy origins, astrophysical black shells can largely be modeled using classical physics, provided that the matter is described by a suitable equation of state. The electromagnetic and gravitational characteristics of these shells will be further examined in the remainder of this article.

\section{Electromagnetic properties of the black shell}\label{sec:em_properties}

\subsection{The black shell is black}\label{sec:shell_is_black}

The defining property of a black hole is that it is black. Any black hole mimicker, such as a black shell, should reproduce this property at least to a reasonable extent. 
In this section we will discover that blackness can be achieved if the shell is made of a material with relative permittivity $ϵ_r$ and relative permeability $μ_r$, such that $ϵ=ϵ_r ϵ_0$ and $μ=μ_r μ_0$, with $ϵ_r=μ_r \gg 1$. Given that the shell is carrying a large number of degrees of freedom, on order of $R^2/l_4^2$, it is reasonable to expect that their response to electric and magnetic fields can be strong, if these degrees of freedom are associated with electric and magnetic dipoles. We will not make any attempt to estimate these values, but plan to return to this question in the future. As we will see, as long as they are sufficiently big, physically measurable quantities will not depend on their exact values.
The speed of light within the material that makes up the thin shell will then be much smaller than the speed of light in vacuum:
\begin{equation}\label{eq:cs}
    c_s \coloneqq \frac{1}{\sqrt{\mu \epsilon}} \ll c
\end{equation}
For this reason, we will refer to the region within the shell as the {\it black domain}.\footnote{This phrase is borrowed from ``Death's end'' by Cixin  Liu, the third book in the trilogy starting with ``The three body problem''. In the book, \emph{black domain} is a region of space where light slows to a standstill.}

Maxwell's equations with $\vec{D}=\epsilon \vec{E}$ and $\vec{B}=\mu \vec{H}$, are given by
\begin{equation}
    \vec{\nabla} \times \vec{H} = \vec{J} + \frac{d\vec{D}}{dt} ,
\end{equation}
where $\vec{J}= \sigma \vec{E}$, and
\begin{equation}
    \vec{\nabla} \times \vec{E} = -\frac{d\vec{B}}{dt} .
\end{equation}
The solution for an electromagnetic wave travelling within the thin shell is proportional to $e^{- \kappa z - i k z + i \omega t}$, where
\begin{equation}
    k= \omega \sqrt{\frac{\epsilon \mu}{2}}\left( \sqrt{1+\left(\frac{\sigma}{\epsilon \omega}\right)^2} +1 \right) ^{1/2}
    \xrightarrow{\lim {σ \ll ϵ ω}}  \omega \sqrt{\mu\epsilon}  ,
\end{equation}
verifying the expected expression for the speed of light in the medium. 

We also find an exponential decay governed by 
\begin{equation}
    \kappa = \omega \sqrt{\frac{\epsilon \mu}{2}}\left( \sqrt{1+\left(\frac{\sigma}{\epsilon \omega}\right)^2} -1 \right) ^{1/2}
    \xrightarrow{\lim {σ \ll ϵ ω}}   \sqrt{\frac{\mu}{\epsilon}} \frac{\sigma}{2}\,.
\end{equation}
The length scale associated to this decay can be identified with a characteristic \emph{skin depth}
\begin{equation} \label{eq:skin_depth}
    δ_s \coloneqq \frac{2}{σ}\sqrt{\frac{ϵ}{μ}},
\end{equation} 
to which fields can penetrate into the shell before being damped by an e-fold. The effective resistance associated with currents flowing within this slice is proportional to $\sigma \delta_s \sim \sqrt{ϵ/μ} =\sqrt{ϵ_0/μ_0}$, which is independent of $\sigma$ and identical to the impedance of the vacuum. This is exactly the same value as in the case of the black hole horizon. Provided that the skin depth is much smaller than the thickness of the shell, $\delta_s \ll \delta a$, the waves will be absorbed quickly and fully, long before they reach the other side of the thin shell.

Using the Fresnel equations, ignoring conductivity,  one notes that any incident wave exactly normal to the surface will be perfectly transmitted without any reflection if  $\epsilon_r = \mu_r$. Taking conductivity into account, there will only be a very small, imaginary, reflection coefficient proportional to $σ/(ϵω) \ll 1$. This is promising for a black shell.

In the case of a wave coming in at an angle, Snell's law tells us that the wave will be refracted towards the normal, and, given the high refractive index within the shell, continue on the inside almost exactly along the normal. According to the Fresnel equations, there will now also exist a reflected wave without any suppression. However, the Fresnel equations assume a sharp boundary with $\epsilon$ and $\mu$ behaving as step functions. In any physical system, this cannot be true. There must be a gradient in the permittivity and permeability such that they increase from the vacuum value to the extreme values of the black domain over a finite width. In the presence of a finite gradient, the refracted wave will not take a sharp turn, but instead smoothly change its direction along a geodesic in the medium. Furthermore, any reflection will be suppressed if the wavelength of the light is much smaller than the width of the step. In our case the wavelength of light will decrease the further into the medium the wave penetrates. Given that $c_s =1/\sqrt{\epsilon \mu}$ becomes very small, there will be an enormous pile up of waves across the step. For simplicity, we assume that the width of the step is much smaller than the skin depth, $\delta_w \ll \delta_s$. In order for reflection to be suppressed we then need
\begin{equation}
    c_s/ \omega \ll \delta_w \ll \delta_s\,.
\end{equation}
This is also compatible with our assumption that $σ/(ϵω) \ll 1$. Hence, any reflection due to either the step or conductivity is suppressed. One may note that suppression is strictly speaking needed only for wavelengths $ \sim c/\omega \ll a$, where $a$ is the radius of the shell. This implies that $\sigma a/ \left(\epsilon c\right)$ can be of order one. We conclude that all waves with a wavelength much smaller than the Schwarzschild radius will be fully absorbed, and the black shell will look perfectly black.
\emph{We expect these considerations to be relevant for any black hole mimicker.}

Let us see how this works in more detail. 
For the purpose of this problem, it suffices to zoom in on the shell and look at a tiny section that is effectively a flat slab of material with a graded refractive index. We will work in Cartesian coordinates, and put the boundary at $z=0$, with the positive $z$-axis pointing towards the black domain. 
For definiteness, we choose the polarization of the incoming light such that $E_y(x,z)$, $H_x(x,z)$ and $H_z(x,z)$ are nonzero. For normal incidence, $H_z=0$. We put $\mu = \mu_0 \alpha (z)$ and $\epsilon = \epsilon_0 \alpha (z)$, with $\alpha (z)$ representing a step of width $\delta _w$ around $z=0$, such that $\alpha \rightarrow 1$ as $z \rightarrow - \infty$, and $\alpha \rightarrow \epsilon_r=\mu_r$ as $z \rightarrow  + \infty$.
Taking the electric and magnetic fields to be $\vec{E} = E_y(x,z)e^{i ω t} \hat{y}$, $\vec{H} = H_x(x,z)e^{i ω t} \hat{x} + H_z (x,z)e^{i ω t} \hat{z}$, Maxwell's equations can be written as a set of coupled, driven wave equations:
\begin{align}
    ∇² E_y+ \omega^2 \mu \epsilon E_y - i \omega \frac{d \mu}{dz} H_x&=0, \nonumber \\
    ∇² H_x+ \omega^2 \mu \epsilon H_x - i \omega \frac{d \epsilon}{dz} E_y + \frac{1}{μ}\frac{dμ}{dz}\frac{∂H_z}{∂x}&=0,\\
    ∇² H_z+ \omega^2 \mu \epsilon H_z + \frac{∂}{∂z}\left( \frac{1}{\mu}\frac{d \mu}{dz} H_z\right)&=0, \nonumber
\end{align}
where $∇² \coloneqq \left( ∂_x² + ∂_z ²\right)$.
In case of normal incidence, where $k_x=0$, the third equation is trivially solved by $H_z=0$, while the first two equations decouple and are solved by $E_y \sim e^{-i k_z \int^z \alpha dz}$, with $k_z = \omega/c$ and $H_x=-\sqrt{ϵ_0/μ_0} E_y$.
This is a purely ingoing wave with varying momentum but no reflection, in agreement with Fresnel. If the light comes in at an angle, with $k_x \neq 0$, it is easiest to solve the third equation.
Writing $H_z(x,z) = e^{-i k_x x}f(z)/\sqrt{\mu(z)}$, the last equation above becomes (here and below, primes denote derivatives with respect to $z$)
\begin{equation} \label{eq:difff}
    f'' + \left( \left( \frac{\omega}{c}\right)^2 \alpha^2 - k_x^2 + \frac{α''}{2 \alpha}-\frac{3\left(α'\right)²}{4 \alpha^2} \right)f =0
\end{equation}
\begin{figure}
    \centering
    \def\svgwidth{0.9\linewidth}
    %% Creator: Inkscape 1.2.2 (b0a84865, 2022-12-01), www.inkscape.org
%% Accompanies image file 'graded.pdf' (pdf, eps, ps)
\begingroup%
  \makeatletter%
  \providecommand\color[2][]{%
    \errmessage{(Inkscape) Color is used for the text in Inkscape, but the package 'color.sty' is not loaded}%
    \renewcommand\color[2][]{}%
  }%
  \providecommand\transparent[1]{%
    \errmessage{(Inkscape) Transparency is used (non-zero) for the text in Inkscape, but the package 'transparent.sty' is not loaded}%
    \renewcommand\transparent[1]{}%
  }%
  \providecommand\rotatebox[2]{#2}%
  \newcommand*\fsize{\dimexpr\f@size pt\relax}%
  \newcommand*\lineheight[1]{\fontsize{\fsize}{#1\fsize}\selectfont}%
  \ifx\svgwidth\undefined%
    \setlength{\unitlength}{324.97160664bp}%
    \ifx\svgscale\undefined%
      \relax%
    \else%
      \setlength{\unitlength}{\unitlength * \real{\svgscale}}%
    \fi%
  \else%
    \setlength{\unitlength}{\svgwidth}%
  \fi%
  \global\let\svgwidth\undefined%
  \global\let\svgscale\undefined%
  \makeatother%
  \begin{picture}(1,0.62051498)%
    \lineheight{1}%
    \setlength\tabcolsep{0pt}%
    \put(0,0){\includegraphics[width=\unitlength,page=1]{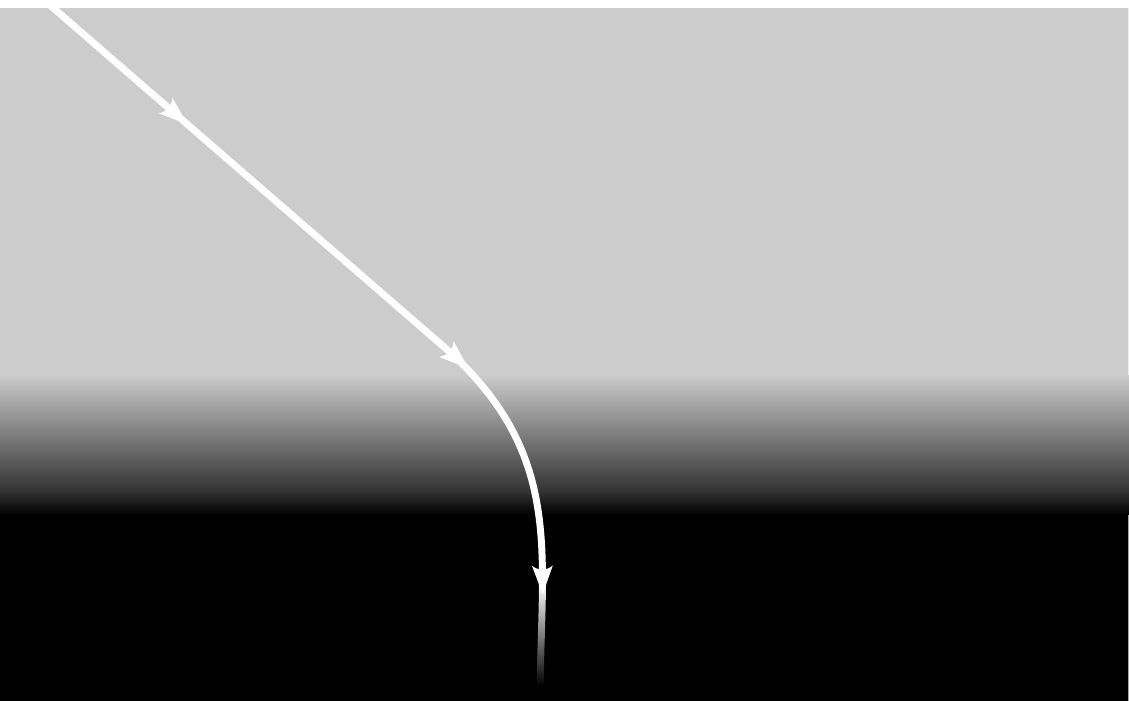}}%
    \put(0.64054465,0.04161327){\color[rgb]{1,1,1}\makebox(0,0)[lt]{\lineheight{1.25}\smash{\begin{tabular}[t]{l}$μ=μ_rμ_0, ϵ=ϵ_rϵ_0$\end{tabular}}}}%
    \put(0.67197162,0.49004868){\color[rgb]{0,0,0}\makebox(0,0)[lt]{\lineheight{1.25}\smash{\begin{tabular}[t]{l}$μ=μ_0, ϵ=ϵ_0$\end{tabular}}}}%
  \end{picture}%
\endgroup%

    \caption{A sketch showing that a ray of light incident at an angle to the black shell bends smoothly towards the normal and continues along the normal thereafter, until it gets completely absorbed. The refractive index of the medium, which increases smoothly from its vacuum value outside the shell to a large value in the shell, is shown as a graded background. The dimming of the light ray as it passes through the shell represents a reduction in its intensity until it is absorbed into the shell.}
    \label{fig:black domain}
\end{figure}
Through Maxwell's equations, the solutions for $H_z$ will also generate solutions for the other components $E_y$ and $H_x$. The solution is uniquely specified by the condition that it is an incoming wave at an angle. In the limit of normal incidence, $k_x=0$, the exact solution is 
\begin{equation} \label{eq:solf}
f=\frac{1}{\sqrt{\alpha}}e^{-i k_z \int^z \alpha dz}\,,
\end{equation}
which implies zero reflection compatible with the Fresnel result. The limit is, however, degenerate since the amplitude of $H_z$ needs to be put to zero for normal incidence. Turning on $k_x$, we can no longer easily find an exact solution. (Note that $k_x^{\rm max} = \omega/c$, corresponding to a grazing ray of light). To better understand the physics, we change coordinates to $d\zeta= \alpha (z)dz$. In this new coordinate, the wavelength of light becomes essentially constant within the black domain. We then introduce $g(z) = \sqrt{\alpha (z)}f(z)$, to find the very simple equation
\begin{equation}
      -\frac{d^2 g}{d\zeta^2} + V(\zeta ) g =0\,,
\end{equation}
with the potential
\begin{equation} \label{eq: simple V}
    V(z) = \frac{k_x^2}{\alpha(z(\zeta))^2}-\left(\frac{\omega}{c}\right)^2
\end{equation}
such that we can think of the problem as a one-dimensional Schrödinger equation for a particle passing over a barrier. We note that for large $|ζ|$ (i.e. large $|z|$, where $α(z)$ becomes effectively constant) the potential goes to a constant according to
\begin{eqnarray}
   V \rightarrow   k_x^2-\left(\frac{\omega}{c}\right)^2 \,\,\, {\rm for}\,\,\, \zeta \rightarrow - \infty\\
   V \rightarrow  \frac{k_x^2}{\alpha_r^2}-\left(\frac{\omega}{c}\right)^2\,\,\, {\rm for} \,\,\,\zeta \rightarrow + \infty
\end{eqnarray}
where $\alpha_r=\epsilon_r=\mu_r$ is large. 
Since $\alpha$ is monotonic in $z$ (and $\zeta$), the potential $V$ is a smooth and monotonic function that goes from the negative value $k_x^2-\left(ω/c\right)^2$ far from the black domain, to the even more negative value $(k_x/α_r)² -(ω/c)²$ deep inside the black domain.

As we have seen, reflection is identically zero for $k_x=0$. This is obvious from (\ref{eq: simple V}), since the potential is constant everywhere. For $k_x\neq 0$ this is no longer true, but any reflection will still be heavily suppressed. In the original coordinate $z$, the step has width $\delta_w$, while in the new coordinate $\zeta$ the width is rescaled to $\alpha_r \delta_w$. Although $\delta_w$ is supposed to be much {\it smaller} than any realistic wave length, $\alpha_r \delta_w$ is assumed to be much {\it larger}. This implies that we are in an essentially classical regime, and the reflection can be neglected for all values of $k_x$. This is sketched in \cref{fig:black domain}.

What happens is that the graded boundary of the black domain mimics space-time just outside of the horizon of a black hole. That is, light rays are bent so that they travel along the normal. This is compatible with the well-known electromagnetic boundary condition on the horizon \cite{damour1978black,Thorne:1986iy,znajek1978electric}, which says that the normal components $\vec{E}_n$ and $\vec{B}_n$, as well as the parallel components, $\vec{E}_H$ and $\vec{B}_H$, as measured by a freely falling observer are finite. In addition 
\begin{equation}
    \vec{E}_H = \vec{n} \times \vec{B}_H \quad , \quad  \vec{B}_H = -\vec{n} \times \vec{E}_H \, ,
\end{equation}
which implies that all electromagnetic fields near the horizon, look like incoming waves along the normal. This can be understood as a consequence of a diverging Lorentzboost.\footnote{The similarity between Maxwell's equations with nontrivial permittivity and permeability, keeping the impedance $\sqrt{ϵ/μ}$ constant, and electromagnetism in a curved spacetime (in particularly close to a horizon) has previously been observed in \cite{Reznik:1997ag}.} The fully developed black domain works more like the horizon itself--- including its electric properties as we will see later on.

\subsection{The black shell has no hair}
\label{sec:no_hair}

As an attempt to distinguish a black shell from a black hole, one may wonder about the fate of a charge that falls onto the stationary black shell. Had it been a black hole, the information about the angular coordinates where the point charge is dropped in will eventually be lost to an external observer after the charge passes the horizon, and any measurement performed will then see electric field lines emanating radially from the spherical horizon of the Schwarzschild black hole. This is a consequence of the black hole \emph{no-hair} theorem. Since the black shell lacks a horizon, one does not naively expect such a no-hair effect. On the other hand, the black shell has a large number of degrees of freedom by construction, and one may expect the charge to somehow ``dissolve'' into these degrees of freedom, with the whole black shell carrying the charge. But since charge is quantized, if the no-hair effect is to be redeemed, dissolving can not be the correct solution.

Another possibility is that the shell simply has a non-zero conductivity, and that charges redistribute themselves so that the entire shell is at the same electric potential. This is a simple way to satisfy the no-hair theorem. As we have already seen, we do need a non-zero conductivity for other reasons, but this cannot on its own explain no hair. Due to the large permittivity, that we needed for the black shell to be black, the electric field sourced by a charge within the black domain will be extremely weak (as we will show in detail below), and any current much to weak to realize the no-hair theorem fast enough.  

Luckily, as we will now show, the large permittivity in itself implies no hair. But this is not achieved through free charges that redistribute due to currents, but through bound charges that appear at the surface. To figure out what is going on, we consider a point charge embedded within the shell and compute the electric field that results. Since the shell sits at $r=9M/4$, and we are interested in the effect of the point charge within a stellar mass or larger shell; we will, for the moment, disregard the spacetime curvature and work in flat space. The qualitative picture will be the same. In this limit, this is simply a problem in classical electrostatics. The general case of a point charge embedded both within and inside a dielectric shell was discussed in \cite{2022arXiv220807706M}. In the following, we will reproduce their results for the case that is of interest to us.

To solve for the electric field, we need to solve the Poisson equation for the scalar potential in three regions: inside, within, and outside the shell. While there are no sources inside and outside the shell, the point charge within the shell provides a source term proportional to the charge.
\begin{figure*}
    \centering
    \subfigure[$\vec{E}$]{
    \includegraphics[width=0.45\linewidth]{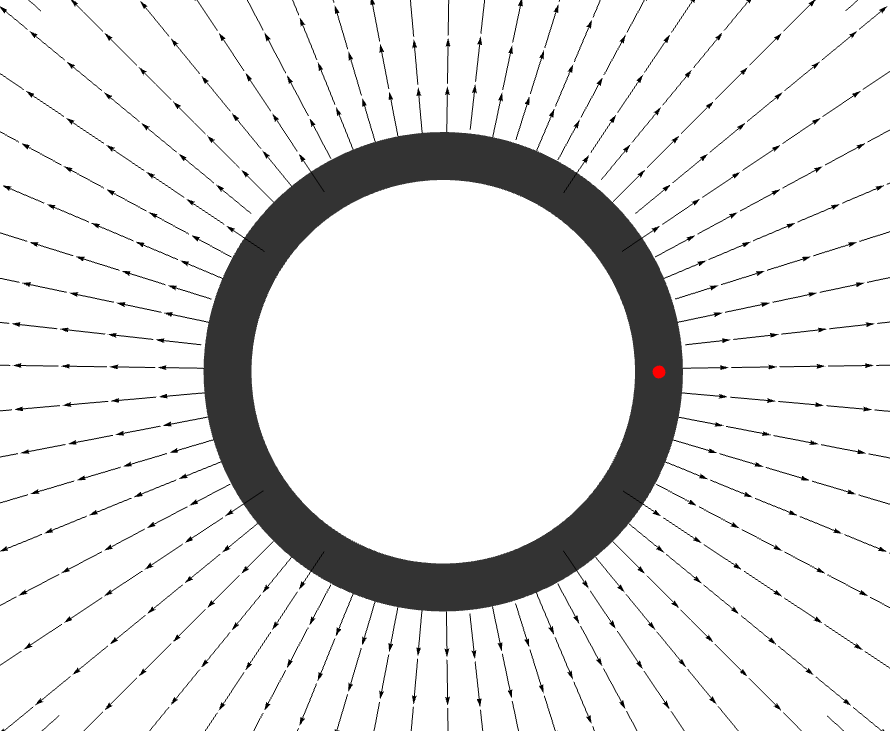}
    \label{fig:E_no_hair}
    }
    \hfill
    \subfigure[$\vec{D}$]{
    \includegraphics[width=0.45\linewidth]{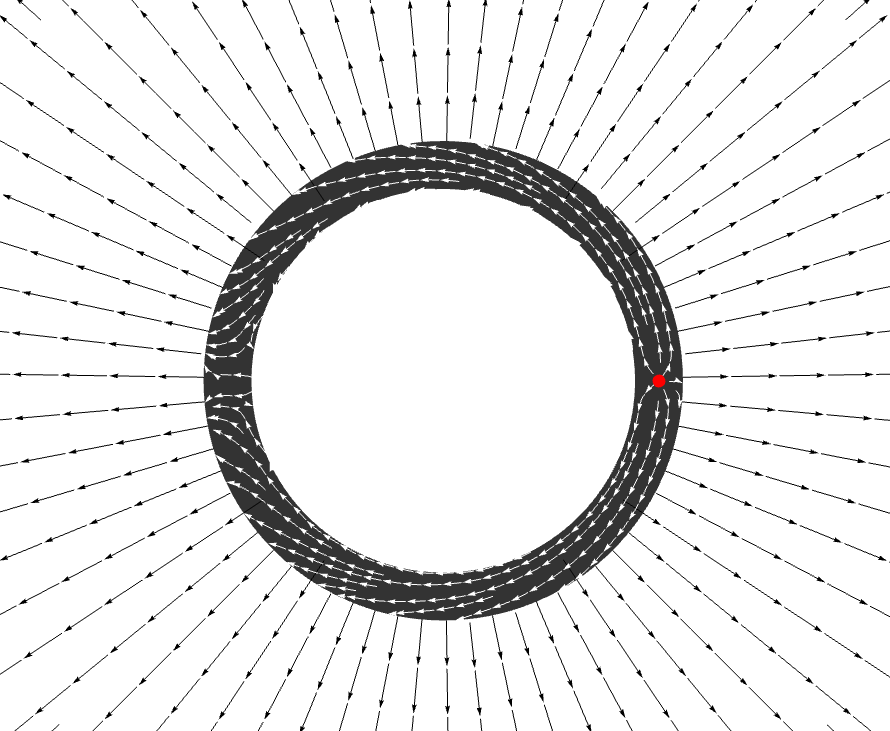}
    \label{fig:D_no_hair}
    }
    \caption{Plots showing the electric field $\vec{E}$ and the displacement field $\vec{D}$ for a point charge (in red) embedded within a shell (shown in black) of infinitely large relative permittivity $ϵ→∞$. The location of the charged particle within the shell is completely shielded from an external observer, giving a no-hair property for the black shell.}
    \label{fig:no-hair}
\end{figure*}
Let us consider a shell of inner radius $a$, outer radius $b$ containing a point charge $q$ at a radial distance $s$ (and $θ=0$), where $a<s<b$. We now make the following ansatz for the scalar potential $Φ$ (assuming regularity at the origin and at infinity):
\begin{align}
    \phiin &= \sum_{n=0}^{∞} A_n r^n P_n,\nonumber\\
    \phishell &=
         \frac{q}{ϵ \left| \vec{r}-\vec{s}\right|}+\sum_{n=0}^{∞} \left(B_n r^n +C_n r^{-n-1} \right)P_n,\\
    \phiout &= \sum_{n=0}^{∞} D_n r^{-n-1} P_n.\nonumber
\end{align}
${\left| \vec{r}-\vec{s}\right|}^{-1}$ accounts for the point charge source within the shell, while the other terms are Green's functions for the Laplace equation. $P_n=P_n(\theta)$ are the Legendre polynomials. The scalar potential on the shell can be rewritten using the standard expansion from \cite{Jackson:1998nia},
\begin{equation}
    \frac{1}{\left| \vec{r}-\vec{s}\right|} = \left\{ 
    \begin{array}{lr}
        \displaystyle \sum_{n=0}^{∞} r^n s^{-n-1} P_n       &   (a<r<s)\,,  \\
        \displaystyle \sum_{n=0}^{∞} s^n r^{-n-1} P_n        &   (s<r<b)\,.
    \end{array}
    \right.
\end{equation}
Boundary conditions are given by:
\begin{align}
    \left(\frac{∂ \phiin}{∂r} = ϵ\frac{∂ \phishell}{∂r} \right)\bigg|_{r=a},&
    \left(\frac{∂ \phiin}{∂θ}= \frac{∂ \phishell}{∂θ}\right) \bigg|_{r=a}\,,\\
    \left(\frac{∂ \phiout}{∂r} = ϵ\frac{∂ \phishell}{∂r} \right)\bigg|_{r=b},&
    \left(\frac{∂ \phiout}{∂θ} = \frac{∂ \phishell}{∂θ}\right) \bigg|_{r=b}.
\end{align}
A solution to this system of equations is given by (after correcting a minor misprint in \cite{2022arXiv220807706M})
\begin{widetext}
\begin{align}
    A_n &= (2n+1)\frac{q}{E_ns^{n+1}} \left[(n+1)(ϵ -1)
   \left(\frac{s}{b}\right)^{2 n+1}+n(ϵ +1)+1\right],\nonumber\\
   B_n &= \left(1-\frac{1}{ϵ}\right) (n+1) \frac{q}{E_ns^{n+1}} \left[n(ϵ -1)
   \left(\frac{a}{b}\right)^{2 n+1}+(ϵ +ϵ  n+n)
   \left(\frac{s}{b}\right)^{2 n+1}\right],\nonumber\\
   C_n &= \left(1-\frac{1}{ϵ}\right) n \frac{q a^{2 n+1}}{E_n s^{n+1}} \left[(ϵ 
   n+n+1)+(n+1)(ϵ -1)
   \left(\frac{s}{b}\right)^{2n+1}\right],\\
   D_n &= (2 n+1) \frac{q a^{2n+1}}{E_ns^{n+1}} \left[n(ϵ -1)+(ϵ +ϵ  n+n) \left(\frac{s}{a}\right)^{2n+1}\right],\nonumber\\
   E_n &= (ϵ  n+n+1) (ϵ +ϵ  n+n)-n
   (n+1) (ϵ -1)^2 \left(\frac{a}{b}\right)^{2 n+1}.\nonumber
\end{align}
\end{widetext}
Consider the limit in which the shell is thin, $δa/a \ll 1$, where $δa \coloneqq b - a$ is the thickness of the shell, and the shell has a large relative permittivity $ϵ_r→∞$. If the permittivity goes to infinity faster than the thickness goes to zero, i.e. $Δ \coloneqq ϵ_r \, δa/a \gg 1$, the electric field ($\vec{E}=-\vec{∇}Φ$) inside, and within the shell vanishes, while it points radially outward outside the shell:\begin{equation}\label{eq:no_hair}
    \lim_{Δ→∞} E_{\textrm{in}} = 
    \lim_{Δ→∞} E_{\textrm{shell}} = 0,
    \lim_{Δ→∞} E_{\textrm{out}} = \frac{q}{r²}.
\end{equation}

Remarkably, this hides the position of the charge within the shell from an external observer, recovering a no-hair property for the shell.\footnote{For the shell to have no electromagnetic hair, it needs to have $ϵ_r \, δa/a \gg 1$. 
For the classical analysis in this article, this is a well-motivated assumption, and we would like to understand its stringy origin in a future work.}
Within the shell, the displacement vector ($\vec{D}=-ϵ\vec{∇}Φ$) is nontrivial in the limit $ϵ→∞$ and shows that the field lines from the charge bend along the shell. The fields $\vec{E}$ and $\vec{D}$ are shown in \cref{fig:no-hair}. It is intriguing that the same properties of the shell that lead to no-hair also account for why the shell is black. As explained at the end of the last section, this is presumably due to the outer layer of black domain (which refracts the field lines in both cases) acting in a way that is very similar to curved spacetime just outside of an horizon. 

The reason to expect large $\epsilon$ is, as already argued, related to the huge number of degrees of freedom of the gas on top of the shell. If each degree of freedom can induce a dipole moment, the relative permittivity and permeability will be proportional to the number density, which is large. The same argument applies to permeability $\mu$. Since we expect the black shell to be democratic with respect to electric and magnetic fields, we expect $\epsilon_r = \mu _r$ both large. In other words, the impedance will be the same as that of the vacuum. While we are making these arguments in the context of the string inspired black shell model, \emph{they should apply to any model that attempts to explain the origin of BH entropy.}

In the above, we ignored the conductivity of the shell for the simple reason that the electric field for this particular configuration was more or less vanishing. In the case of an incoming electromagnetic wave, the electric field within the black shell is non-vanishing, and the conductivity is crucial for the wave to be absorbed. In the next sections, we will study currents flowing through the shell, where again conductivity will be important. We will first consider a source exterior to the shell, and then consider the shell itself as a generator.

\subsection{The black shell as a conductor}\label{sec:conductor}

\begin{figure}
    \centering
    \def\svgwidth{0.8\linewidth}
    %% Creator: Inkscape 1.2.2 (b0a84865, 2022-12-01), www.inkscape.org
%% Accompanies image file 'resistance_black.pdf' (pdf, eps, ps)
\begingroup%
  \makeatletter%
  \providecommand\color[2][]{%
    \errmessage{(Inkscape) Color is used for the text in Inkscape, but the package 'color.sty' is not loaded}%
    \renewcommand\color[2][]{}%
  }%
  \providecommand\transparent[1]{%
    \errmessage{(Inkscape) Transparency is used (non-zero) for the text in Inkscape, but the package 'transparent.sty' is not loaded}%
    \renewcommand\transparent[1]{}%
  }%
  \providecommand\rotatebox[2]{#2}%
  \newcommand*\fsize{\dimexpr\f@size pt\relax}%
  \newcommand*\lineheight[1]{\fontsize{\fsize}{#1\fsize}\selectfont}%
  \ifx\svgwidth\undefined%
    \setlength{\unitlength}{323.7229691bp}%
    \ifx\svgscale\undefined%
      \relax%
    \else%
      \setlength{\unitlength}{\unitlength * \real{\svgscale}}%
    \fi%
  \else%
    \setlength{\unitlength}{\svgwidth}%
  \fi%
  \global\let\svgwidth\undefined%
  \global\let\svgscale\undefined%
  \makeatother%
  \begin{picture}(1,0.90805117)%
    \lineheight{1}%
    \setlength\tabcolsep{0pt}%
    \put(0,0){\includegraphics[width=\unitlength,page=1]{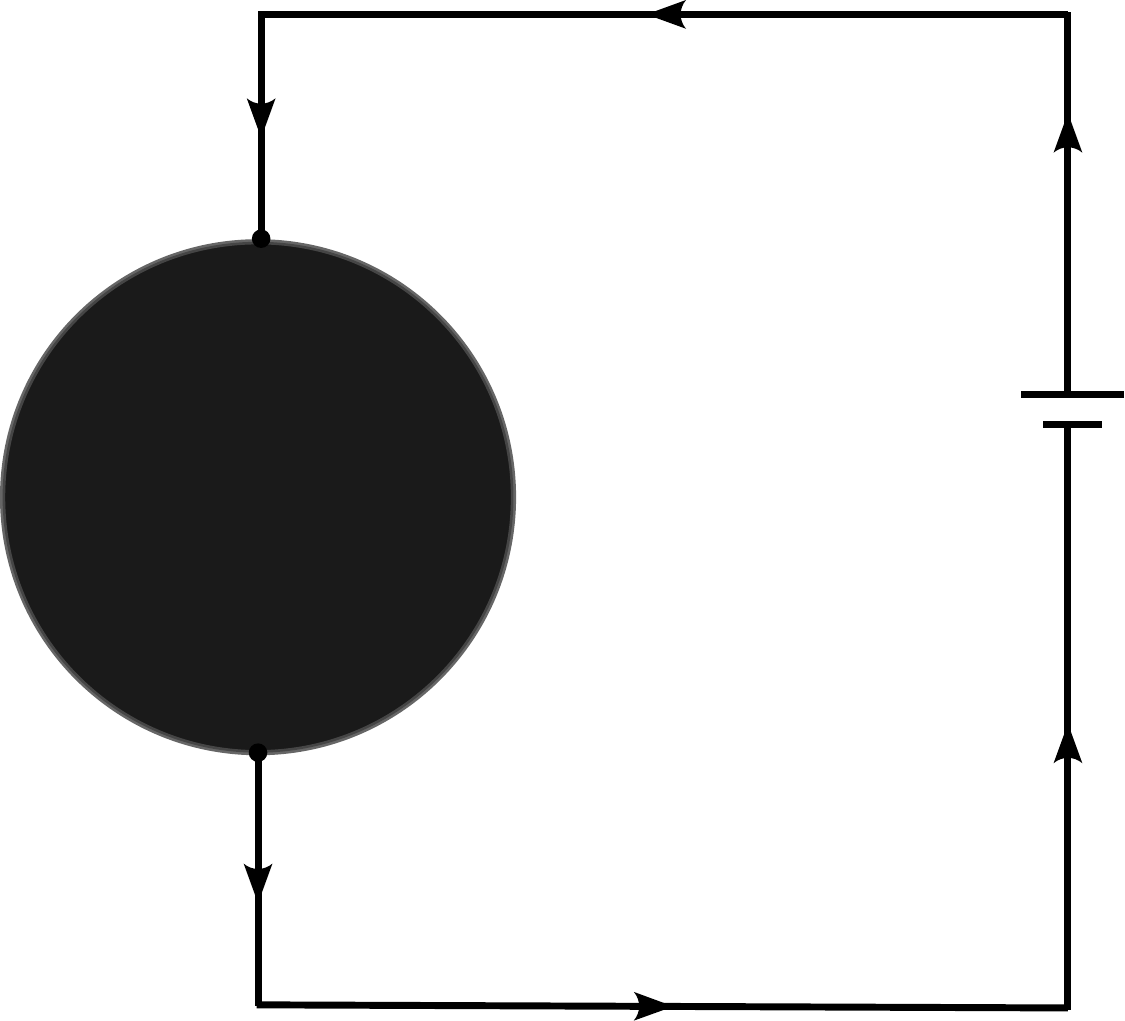}}%
    \put(0.81749621,0.52544275){\color[rgb]{0,0,0}\transparent{0.85882401}\makebox(0,0)[lt]{\lineheight{1.25}\smash{\begin{tabular}[t]{l}$V$\end{tabular}}}}%
    \put(0,0){\includegraphics[width=\unitlength,page=2]{resistance_black.pdf}}%
  \end{picture}%
\endgroup%

    \caption{A schematic of a black shell as a resistor in a circuit with an external battery.}
    \label{fig:resistance}
\end{figure}

\begin{figure*}
    \centering
    \subfigure[$\vec{B}$]{
    \includegraphics[width=0.45\linewidth]{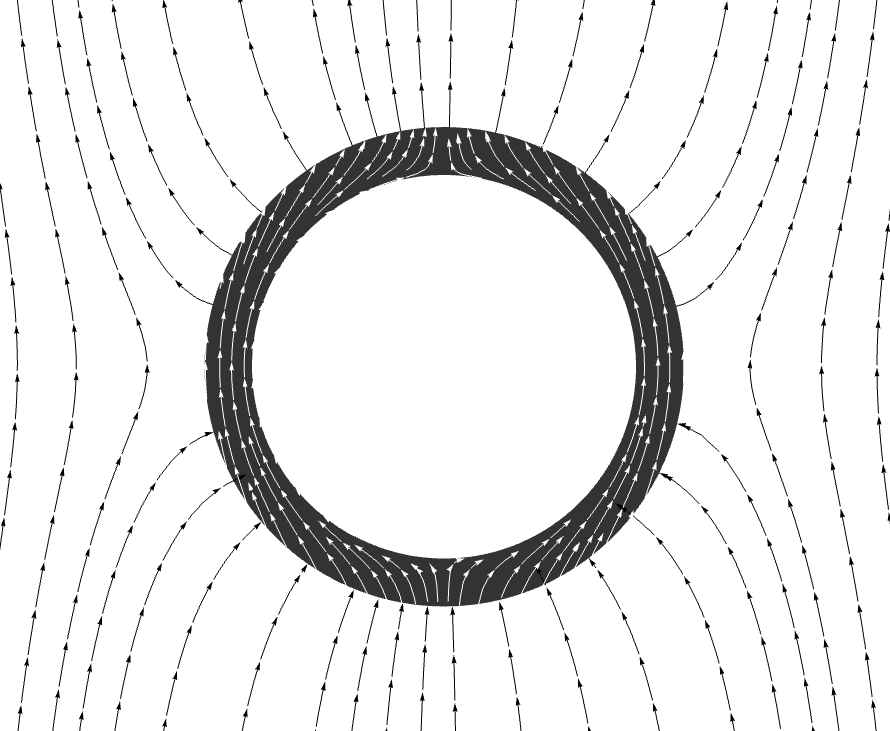}
    \label{fig:B_stationary}
    }
    \hfill
    \subfigure[]{
    \def\svgwidth{0.48\linewidth}
    %% Creator: Inkscape 1.2.2 (b0a84865, 2022-12-01), www.inkscape.org
%% Accompanies image file 'generator_black.pdf' (pdf, eps, ps)
\begingroup%
  \makeatletter%
  \providecommand\color[2][]{%
    \errmessage{(Inkscape) Color is used for the text in Inkscape, but the package 'color.sty' is not loaded}%
    \renewcommand\color[2][]{}%
  }%
  \providecommand\transparent[1]{%
    \errmessage{(Inkscape) Transparency is used (non-zero) for the text in Inkscape, but the package 'transparent.sty' is not loaded}%
    \renewcommand\transparent[1]{}%
  }%
  \providecommand\rotatebox[2]{#2}%
  \newcommand*\fsize{\dimexpr\f@size pt\relax}%
  \newcommand*\lineheight[1]{\fontsize{\fsize}{#1\fsize}\selectfont}%
  \ifx\svgwidth\undefined%
    \setlength{\unitlength}{468.04670784bp}%
    \ifx\svgscale\undefined%
      \relax%
    \else%
      \setlength{\unitlength}{\unitlength * \real{\svgscale}}%
    \fi%
  \else%
    \setlength{\unitlength}{\svgwidth}%
  \fi%
  \global\let\svgwidth\undefined%
  \global\let\svgscale\undefined%
  \makeatother%
  \begin{picture}(1,0.9432404)%
    \lineheight{1}%
    \setlength\tabcolsep{0pt}%
    \put(0,0){\includegraphics[width=\unitlength,page=1]{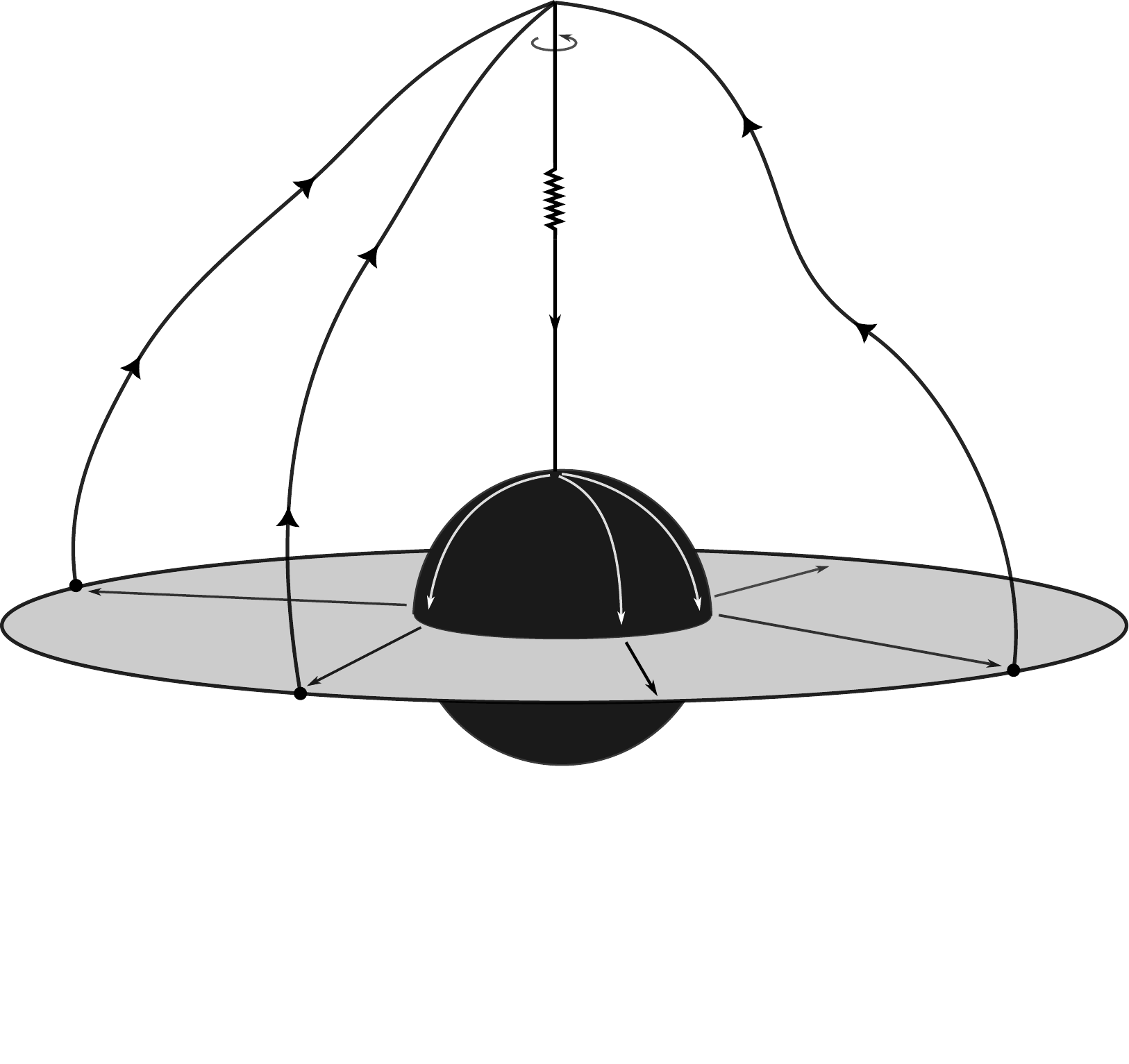}}%
    \put(0.51396668,0.90190599){\color[rgb]{0.2,0.2,0.2}\makebox(0,0)[lt]{\lineheight{1.25}\smash{\begin{tabular}[t]{l}$\Omega$\end{tabular}}}}%
    \put(0.7797313,0.01651901){\color[rgb]{0.2,0.2,0.2}\makebox(0,0)[lt]{\lineheight{1.25}\smash{\begin{tabular}[t]{l}$\vec{B}$\end{tabular}}}}%
    \put(0,0){\includegraphics[width=\unitlength,page=2]{generator_black.pdf}}%
  \end{picture}%
\endgroup%

    \label{fig:emf}
    }
   \caption{(a) Plot showing the magnetic field outside and within the cross-section of a static shell with large $μ,ϵ$ placed in a uniform constant external magnetic field pointing vertically upwards. (b) Schematic of a rotating black shell placed within a circuit in an external magnetic field. It acts like a generator in a closed electric circuit with current entering at the poles and leaving through a conductor in the form of an accretion disc at the equator. There is also an external load.
    }
\end{figure*}

Consider a black shell with a wire connected to the north pole and another wire connected to the south pole; see \cref{fig:resistance}. We then connect the wires to an external voltage source and expect a current to start flowing through the circuit, including the black shell. As in the case of a black hole we do not expect matter to be able to leave the shell. In order for a current to flow, we therefore need, say, electrons entering the shell through one of the wires and positrons entering through the other.\footnote{A scenario already familiar
 in both neutron stars and black holes~\cite{1969ApJ...157..869G,Blandford:1977ds}.}
Maxwell's equations for the conducting shell are simply given by 
\begin{equation}
\begin{aligned}
      \nabla \times E &= -∂_t B\,,  &&\nabla \times H = \sigma E + ∂_t D\,, \\
      \nabla \cdot H &=0\,,  &&\nabla  \cdot E =0\,.
\end{aligned}
\end{equation}
Here we have used the constitutive equations for the shell:
\begin{equation}
D = \epsilon E ,\quad B = \mu H, \quad J = \sigma E.
\end{equation}
We note that the electric field is source-free, $∇\cdot E =0$, everywhere, except where the external wires are connected. In the static case, the fields will penetrate the shell in a way similar to \cref{fig:B_stationary}. However, in a physical situation there will always be some time dependence, with the importance of such a time dependence amplified by $\mu$ and $\epsilon$. 

Let us examine this in more detail. It is useful to split the fields into components parallel to the equator, labeled by a subscript $||$, and the remainder labeled by a subscript $\perp$ (which is normal to the parallel component):
\begin{align}\label{eq:Eperp_Hparallel}
      \nabla \times E_\perp =-\mu \partial_t H_\parallel, \quad \nabla \times H_\parallel = \sigma E_\perp +\epsilon \partial_t E_\perp, \\
      \nabla \times E_\parallel =-\mu \partial_t H_\perp, \quad \nabla \times H_\perp = \sigma E_\parallel +\epsilon \partial_t E_\parallel\,.
\end{align}
We now consider a situation where $E_\parallel = H_\perp=0$ so that we only have a current flowing in the poloidal direction, generating a magnetic field around the equator. It is convenient to combine the two equations in \eqref{eq:Eperp_Hparallel} into 
\begin{equation}
    \nabla \times (\nabla \times H_\parallel ) +\sigma \mu \partial_t H_\parallel +\epsilon \mu \partial_t^2 H_\parallel = 0\,.
\end{equation}
Using $H_\parallel = h(r,\theta ,t) \hat{\phi}$, we find
\begin{equation}
    -\Delta h +\sigma \mu \partial_t h +\epsilon \mu \partial_t^2 h = 0\,,
\end{equation}
where $\Delta \equiv (1/r) ∂_r²(rh) + (1/r²)∂_θ\left( (1/\sin θ) ∂_θ (h \sin θ) \right)$ is the $\hat{ϕ}$ component of the  {\it vector} Laplacian. We make the ansatz
\begin{equation}
    h(r, \theta , t) = \sum_m \frac{1}{r} A_{\omega,m} (r) B_m (\theta ) e^{i \omega t}\,,
\end{equation}
where $B_m(θ)$ is an eigenfunction of the angular operator
\begin{equation}
    \frac{\partial}{\partial \theta}\left( \frac{1}{\sin{\theta}} \frac{\partial}{\partial \theta} \left( \sin{\theta} B_m ( \theta ) \right)  \right) = -m^2 B_m (\theta )\,,
\end{equation}
and can be written in terms of hypergeometric functions:
\begin{align}
    B_m(θ) &= {}_2F_1\left[ -\frac{1}{4}-λ,-\frac{1}{4}+λ;\frac{1}{2};\cos²θ \right] \csc θ\, c_1 \\
    &\hspace{11pt} + {}_2F_1\left[ \frac{1}{4}-λ,\frac{1}{4}+λ;\frac{3}{2};\cos²θ \right]\cot θ\, c_2 \,,
\end{align}
with $λ \coloneqq \sqrt{1+4m²}/4$. The leads to the radial equation:
\begin{equation}
    \frac{\partial^2 A_{n,m}}{\partial r^2} +\left(k^2-\frac{m^2}{r^2} \right) A_{\omega,m} =0\,.
\end{equation}
where (in the limit $σ \ll ϵω$)
\begin{equation}
    k=\sqrt{\epsilon \mu \omega^2 -\sigma \mu \omega i} \sim
    \sqrt{\epsilon \mu} \omega -i \frac{σ}{2}\sqrt{\frac{\mu}{\epsilon}}\,.
\end{equation}
The physically relevant solution is the one that decays as we go into the black domain. This is proportional to a modified Bessel function of the second kind:
\begin{equation}
    A_{\omega,m}=\sqrt{r} K_{2λ} (-i k r)\,,
\end{equation}
where the imaginary part of $k$ gives the expected exponential fall off. In summary, we find our solution
\begin{equation}
  h(r, \theta , t) =  \sum_m \frac{1}{\sqrt{r}} K_{2λ} ( -i k r)  e^{i \omega t}  B_m (\theta )\,.
\end{equation}
We now use $K_\nu (z) \sim \sqrt{\frac{\pi}{2 z}} e^{-z}$ for large $|z|$ (and $ \arg z < 3π/2$) to get~\footnote{ 
We assume that the skin depth in \eqref{eq:skin_depth} is small compared to the thickness of the shell. This requires a large $σ$, which implies large $|z|$.
}
\begin{equation}
  h(r, \theta , t) \sim  \frac{1}{r} e^{r
  \frac{σ}{2}\sqrt{μ/ϵ}} e^{i \sqrt{ \epsilon \mu} \omega r+ i \omega t} \,.
\end{equation}
The radial component decays exponentially
as $r$ decreases from its maximal value at the boundary of the shell. The length scale associated with this decay is again given by the skin depth $\delta_s$ defined in \eqref{eq:skin_depth}.
Just as in the case of the EM wave, we note that $\sigma$ cancels out in the effective resistance, and we reproduce (at large $\omega$) the value for a black hole. The requirements for this to happen are:
\begin{equation}
    \omega >\frac{\sigma}{\epsilon}, \quad \delta_s \ll \delta a\,,
\end{equation}
implying
\begin{equation}
    \omega \gg \frac{c_s}{\delta a}\,,
\end{equation}
where, as before, $δa$ is the thickness of the shell.
This means that any variation faster than it takes for the slowed down light to cross the thin shell will experience a resistance equal to that of a black hole. By choosing $\epsilon_r=\mu_r$ sufficiently large, this will not lead to any practical difference. In principle, one would expect the resistance to decrease for processes sustained over a very long time. 

The above is the familiar phenomenon of resistance depending on the skin depth. At high frequencies the skin depth is small, and the current is only flowing in a thin slice of the shell. The limiting resistance at high frequencies is finite. At low frequencies, the field is penetrating the entire width and the shell functions as an almost perfect conductor. 

The actual solution for a specific configuration depends on the details of how the wires are attached to the shell. The simplest possibility is when the thin wires have vanishing resistance, and extend into the black domain. The relevant solution then has only $m=0$ turned on, which leads to $E_r =0$ and $E_\theta \neq 0$. The electric field (and the current) is then sourced by the inserted wires all the way down to the skin depth. We then have exactly $K_{1/2} (z) = \sqrt{\frac{\pi}{2 z}}$, and the exact solution is given by
\begin{equation}
  h(r, \theta , t) = \frac{1}{r} e^{r/δ_s} e^{i \sqrt{ \epsilon \mu} \omega r+ i \omega t} \frac{1}{\sin \theta}\,.
\end{equation}
It is also possible to construct solutions where the wires end at the surface, where they appear as localized sources. We will not discuss this here.
 
\section{The black shell as an electric generator producing jets}\label{sec:generator}

Let us now examine the impact of introducing a rotating hollow shell with relative permeability $μ$ and relative permittivity $ϵ$ into a uniform magnetic field. It is reasonable to assume that the external magnetic field will penetrate the shell, resulting in a magnetic field within the black domain. Note, though, that the interior AdS is effectively shielded from any magnetic field. As the shell rotates, the bound charges within it will encounter a Lorenz force. Consequently, this will give rise to a non-uniform distribution of charges within the black domain, leading to a non-trivial electric field.

One key difference between a shell and a solid sphere composed of the same material is that the magnetic field within a solid sphere is directly proportional to the external field, resulting in a constant magnetic field. However, this relationship does not hold true for a shell. As a result, we will approach the problem in two stages.
\begin{enumerate}[label=(\roman*)]
    \item we will first compute the magnetic field within the black domain
    \item using this magnetic field, we will compute the electric field generated by the rotating shell.
\end{enumerate}
\subsection{Magnetic field inside the shell}\label{sec:B_stationary}

Consider a spherical shell of inner radius $a$, and outer radius $b$ of relative permeability $μ$ and relative permittivity $ϵ$ placed in a uniform external magnetic field $\vec{H_0} = H_0 \hat{z}$. Outside the shell (vacuum with $μ_0=1$), $\vec{B}=\vec{H_0} = H_0 \hat{z}$. The scalar field $Φ$ corresponding to a magnetic field $\vec{H}$ is given by $\vec{H} = - \vec{∇} Φ$. Outside the shell, $\vec{∇} Φ = -H_0 \hat{z}$, which gives (in radial coordinates centered at the origin of the shell)
\begin{equation}
    Φ_0 = -H_0 r \cos θ = -H_0 r P_1\,,
\end{equation}
where $P_n$ is the $n$-th Legendre polynomial. To solve Maxwell's equation $\vec{∇}\cdot \vec{B} = 0 = ∇² Φ$, we make the following ansatz for $Φ$ (assuming regularity at the origin and at infinity):
\begin{equation}\label{eq:phi_ansatz_stationary}
    Φ = \left\{ 
    \begin{array}{lr}
         A r P_1            &   (r<a)\,,  \\
         B r P_1 + C P_1/r² &   (a<r<b)\,,\\
         -H_0 r P_1 +D P_1/r²&   (r>b)\,.
    \end{array}
    \right.
\end{equation}
Boundary conditions are given by continuity of the potential $Φ$ and continuity of the radial component of the magnetic field $\vec{B}=μ\vec{H}$ (which follows from $\vec{∇}\cdot \vec{B}$) across the shell at $r=a,b$:
\begin{align}\label{eq:bc_stationary}
    A           &= B + C/a³\,,\nonumber\\
    B + C/b³    &= -H_0 + D/b³\,,\nonumber\\
    A           &= μ (B-2C/a³)\,,\\
    μ(B-2C/b³)  &= -H_0 - 2D/b³\,.\nonumber
\end{align} 
Solving \eqref{eq:phi_ansatz_stationary} with the above boundary conditions gives
\begin{align}\label{eq:phi_sol_stationary}
    A &= -9μ b³\frac{H_0}{E}\,,\nonumber\\
    B &= -3(1+2μ)b³\frac{H_0}{E}\,,\nonumber\\
    C &= 3(1-μ)a³b³\frac{H_0}{E}\,,\\
    D &= (1+μ-2μ²)(b³-a³)b³\frac{H_0}{E}\,,\nonumber
\end{align}
with $E \coloneqq -2a³(1-μ)²+b³(2+5μ+2μ²)$.
In the limit of large $μ,ϵ$, this results in the following magnetic field $\vec{B} = μ \vec{H} = - μ \vec{∇} Φ$:
\begin{align}\label{eq:B_stationary}
    &\vec{B}_{\textrm{in}} = \mathcal{O}\left(1/μ\right)\,,\nonumber\\
    &\vec{B}_{\textrm{shell}} = -\frac{3b³H_0}{2r³\left( b³-a³ \right)} \left[ -\left(r³-a³\right) \cos θ \hat{r}\right.\nonumber\\
    &\hskip 45pt \left. +\left(a³+2r³\right) \sin θ \hat{θ} \right]+\mathcal{O}\left(1/μ\right)\,,\\
    &\vec{B}_{\textrm{out}} = \frac{H_0}{r³} \left[ \left(2b³+r³\right)\cos θ \hat{r} - \left(r³-b³\right) \sin θ \hat{θ} \right]\nonumber\\
    &\hskip 45pt +\mathcal{O}\left(1/μ\right)\,.\nonumber
\end{align}
It should be noted that when the value of $μ$ is large, the interior of the shell is completely protected from the magnetic field. The solution for this scenario is illustrated in \cref{fig:B_stationary}. As expected, the magnetic field within the shell does not remain constant, unlike in the case of a solid sphere. Instead, it exhibits a non-trivial dependence on both $r$ and $θ$.

\subsection{The black shell as a generator}

The black shell acts like a generator in a way very similar to a black hole. Let us assume an external magnetic field similar to the one in the previous section, with a radial component of the magnetic field pointing outwards in the northern hemisphere and inwards in the southern hemisphere. We let the black shell spin with angular velocity $\Omega$ with its axis pointing upwards and connect to it a very long wire that extends upwards from the north pole as illustrated in \cref{fig:emf}. The wire then turns around very far away from the shell and connects back to the equator via sliding contacts mounted on an equatorial disc such that they maintain constant contact with the shell, closing the circuit.
This induces an emf in the closed circuit given by
\begin{equation}
    {\cal E} = \frac{\Omega \Phi}{2 \pi }
\end{equation}
where $\Phi$ is the total magnetic flux entering through the southern hemisphere (and exiting through the northern). This is the same result as for a black hole. A current will flow from the north pole to the equator, generating a magnetic field along the equator. As argued in \cref{sec:conductor}, the current will be flowing in a very thin sheet of the shell. This current shields the interior from the magnetic field pointing in the $\phi$-direction on the outside of the shell.

Let us figure out the structure of the electric and magnetic fields within the thin shell. As discussed above, the radial component of the magnetic field will give rise to a Lorentz force driving a current along the surface of the shell in the direction of $\hat{\theta}$. A bit deeper into the black domain, the magnetic field must turn in the angular direction in order for the fields in the southern and northern hemispheres to be connected. This will give rise to a Lorentz force pushing the current towards the surface. Let us now work out the details. 

To proceed we need to relate the $E$, $D$, $B$ and $H$ fields in the inertial system to the corresponding fields,  $E^*$, $D^*$, $B^*$ and $H^*$ in a frame co-rotating with the shell. We have  the constitutive relations:
\begin{equation}
D^*= \epsilon E^*,\quad B^*= \mu H^*, \quad J^*= \sigma E^*.
\end{equation}
These fields are related to the inertial frame through the Lorentz transformations:
\begin{equation}
\begin{aligned}
    & E_\perp^* = \gamma (E_\perp + v \times B)\,,&& B_\perp ^* = \gamma (B_\perp -  \frac{v}{c^2} \times E)\,,\\
    & H_\perp ^* = \gamma (H_\perp - v \times D)\,,&& D_\perp^* = \gamma (D_\perp + \frac{v}{c^2} \times H)\,.
\end{aligned}
\end{equation}
Note that the component of the current transverse to $v$, has $J_\perp= J_\perp^*$, while the parallel component has $J_\parallel = \gamma J_\parallel^*$. We also note that if there is no charge density in the rest frame of the shell, $\rho^*=0$, there will be one generated in the inertial frame given by $\rho = -\gamma v J_\parallel^*/c^2$.

Maxwell's equations in the frame of the shell are identical to those we used in the previous section:
\begin{equation}
\begin{aligned}
      &\nabla \times E^* =-\mu \partial_t H^*\,,  
      &&\nabla \times H^* = \sigma E^* +\epsilon \partial_t E^*\,, \\
      &\nabla \cdot H^*=0\,, 
      &&\nabla  \cdot E^*=0\,,
\end{aligned}
\end{equation}
where the last relation is due to  $J^*= \sigma E^*$  being conserved everywhere, except where the external wires are connected. If one wants to compare with Maxwell's equations in the inertial frame it is useful to note that $\nabla \times (v \times H^*)=\nabla \times (v \times E^*) =0$. We note, though, that $\nabla \cdot E_\perp = - \nabla \cdot (v \times H^*) \neq 0$, reflecting the cancellation of the Lorentz force pushing the current away from the rotational axis. 

Since the equations for $E^*_\parallel =0$ and $H^*_\perp$ (determining the magnetic field) decouple from $E^*_\perp$ and $H^*_\parallel$ (determining the current), except through boundary conditions, we can use the results from the previous sections. The only difference is how the wires are connected to the shell. For a black shell with an accretion disc, generating jets, we expect something like \cref{fig:emf}. We have thin wires connected at the north and the south pole together with a thin, disc like wire connected to the equator. With the magnetic field as well as the rotation vector pointing upwards, we will obtain a current flowing from the poles to the equator. 

The detailed profile of the solution depends, as discussed previously, on the anatomy of the attached wires. The simplest case is again $m=0$, but now the currents flow along the longitudes from both poles (with $E_r=0$), and are intercepted by the disc shaped conductor that makes sliding contact along the entire equator.

Based on the above analysis, let us now speculate on the structure of the magnetosphere and the mechanism responsible for jet production in the context of a supermassive black shell.

In the accretion disk, gravitational energy is converted into kinetic and thermal energy, which keeps the accreting plasma highly ionized and, therefore, highly conducting. As a result, the magnetic fields within the accretion disk are effectively ``frozen" into the plasma (ideal magnetohydrodynamics far from the shell, with some reconnection of field lines occurring closer to the shell due to Rayleigh-Taylor instabilities). As the plasma slowly accretes toward the shell under the influence of viscosity, it drags the magnetic field lines along, transporting them onto the shell in a manner analogous to that for black holes. Additionally, the inner part of the accretion disk aligns itself with the equatorial plane of the spinning shell (due to the Bardeen-Petterson effect).

As the magnetized plasma accretes onto the shell, closed magnetic field loops are destroyed, and field lines are redistributed, leading to a roughly uniform distribution of field lines threading the shell. All of this is similar to the case of a black hole (see, e.g. Fig. 36 in \cite{Thorne:1986iy}).

The resulting plasma-filled magnetosphere resembles the circuit depicted in \cref{fig:emf}, where the shell’s rotational energy is extracted and directed along the polar regions. This extracted energy can drive the production of jets observed in quasars and active galactic nuclei. While systems with high accretion rates typically have thin disks, supermassive black holes and black shells are expected to accrete at low rates, resulting in thick disks. Although the analysis presented here assumes a thin disk, the results are expected to hold qualitatively for thick disks in slow accretion scenarios characteristic of supermassive systems.

\section{Electromagnetic signatures of black shells}\label{sec:em_observables}

Having discussed electromagnetic properties of the black shell, let us now turn our attention to their observational signatures in the electromagnetic spectrum i.e. images obtained in VLBI experiments like the EHT.

Photons that travel from the object to the observer's camera make up the image on the observer's screen. Given the limited field of view of the camera, most photons emitted near the object never make it to the screen. An efficient way to reconstruct the image on the screen is, therefore, to trace photon trajectories back in time from the screen back towards the source. In the case of a black shell (or a black hole), given enough time, the backward traced photons will either end up at the surface of the black shell (or at the event horizon in the case of a black hole) or escape out of the camera sphere, to infinity. Computing these photon trajectories goes by the name of \emph{ray tracing} and can be efficiently performed by integrating null geodesics in the given spacetime. While this can be done analytically for the Kerr spacetime, the geodesic equations need to be integrated numerically for general spacetimes. 
Let us summarize the equations to solve, and initial conditions below.
\begin{figure}
    \centering
    \def\svgwidth{\linewidth}
    %% Creator: Inkscape 1.2.2 (b0a84865, 2022-12-01), www.inkscape.org
%% Accompanies image file 'new_shadow.pdf' (pdf, eps, ps)
\begingroup%
  \makeatletter%
  \providecommand\color[2][]{%
    \errmessage{(Inkscape) Color is used for the text in Inkscape, but the package 'color.sty' is not loaded}%
    \renewcommand\color[2][]{}%
  }%
  \providecommand\transparent[1]{%
    \errmessage{(Inkscape) Transparency is used (non-zero) for the text in Inkscape, but the package 'transparent.sty' is not loaded}%
    \renewcommand\transparent[1]{}%
  }%
  \providecommand\rotatebox[2]{#2}%
  \newcommand*\fsize{\dimexpr\f@size pt\relax}%
  \newcommand*\lineheight[1]{\fontsize{\fsize}{#1\fsize}\selectfont}%
  \ifx\svgwidth\undefined%
    \setlength{\unitlength}{463.89913947bp}%
    \ifx\svgscale\undefined%
      \relax%
    \else%
      \setlength{\unitlength}{\unitlength * \real{\svgscale}}%
    \fi%
  \else%
    \setlength{\unitlength}{\svgwidth}%
  \fi%
  \global\let\svgwidth\undefined%
  \global\let\svgscale\undefined%
  \makeatother%
  \begin{picture}(1,0.98129499)%
    \lineheight{1}%
    \setlength\tabcolsep{0pt}%
    \put(0,0){\includegraphics[width=\unitlength,page=1]{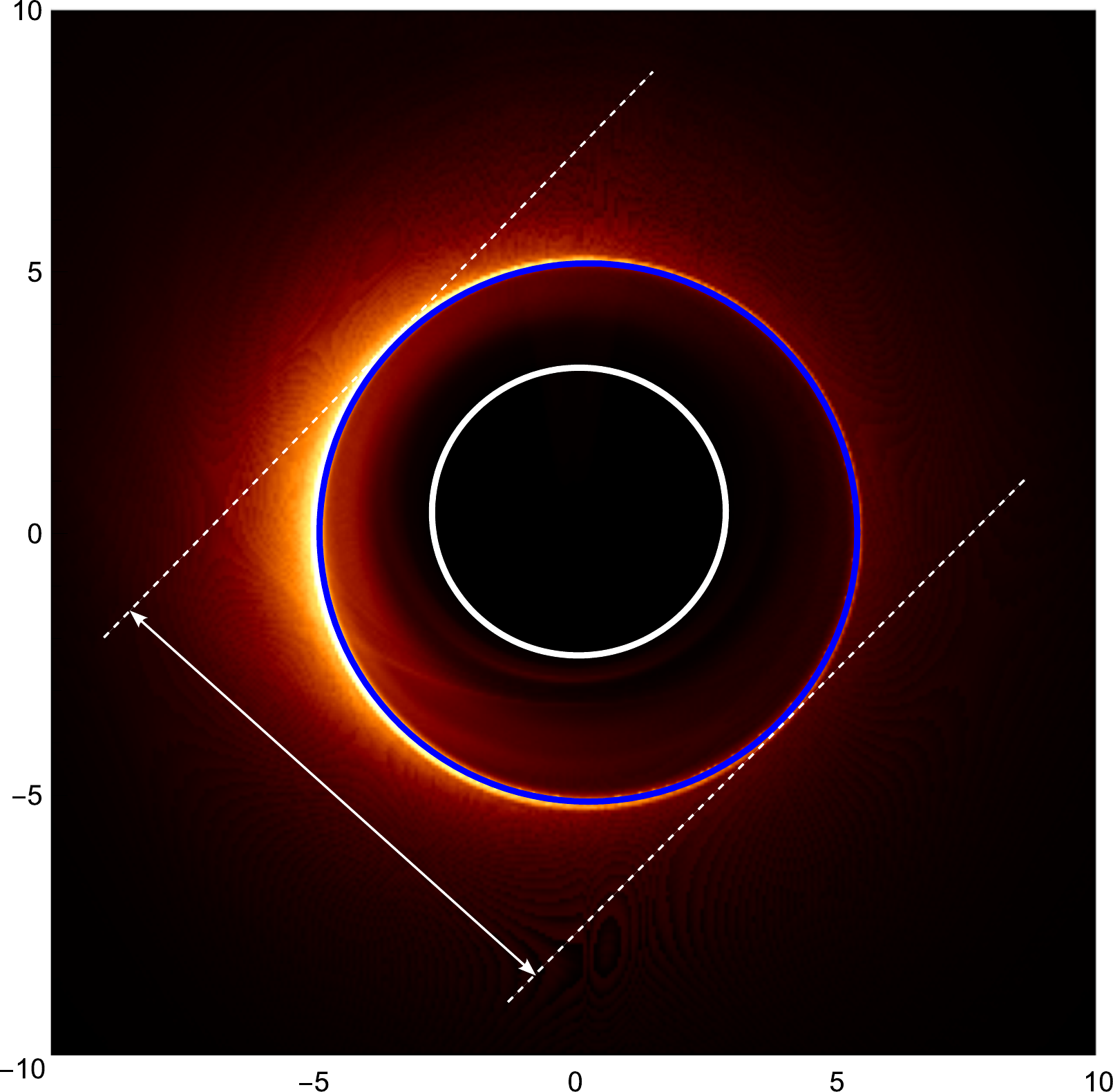}}%
    \put(0.24479247,0.22374593){\color[rgb]{1,1,1}\makebox(0,0)[lt]{\lineheight{1.25}\smash{\begin{tabular}[t]{l}$d(\varphi)$\end{tabular}}}}%
    \put(0,0){\includegraphics[width=\unitlength,page=2]{new_shadow.pdf}}%
    \put(0.49114866,0.15751881){\color[rgb]{1,1,1}\makebox(0,0)[lt]{\lineheight{1.25}\smash{\begin{tabular}[t]{l}$\varphi$\end{tabular}}}}%
    \put(0.4150797,0.51055232){\color[rgb]{1,1,1}\makebox(0,0)[lt]{\lineheight{1.25}\smash{\begin{tabular}[t]{l}inner shadow\end{tabular}}}}%
    \put(0.71583531,0.67829882){\color[rgb]{1,1,1}\makebox(0,0)[lt]{\lineheight{1.25}\smash{\begin{tabular}[t]{l}critical curve\end{tabular}}}}%
  \end{picture}%
\endgroup%

    \caption{Snapshot of simulated synchrotron emission at $230\textrm{GHz}$ (obtained from a GRMHD simulation) of a Kerr black hole with $M=6.5×10^9 M_\odot, a = 0.45$, observed by a camera placed at a distance of $16.9 \textrm{Mpc}$ from the source, and inclined at $17^\circ$ to the spin axis. Superimposed on the image are the critical curve (blue curve), boundary of the inner shadow (white curve), and the projected diameter $d(φ)$. The axes are labeled in units of $M$.}
    \label{fig:curves}
\end{figure}
\subsection{Geodesic equations and initial conditions}
For a stationary, axisymmetric metric in Boyer-Lindquist coordinates
\begin{equation}
    \d s² = g_{tt} \d t² + g_{rr} \d r² + g_{θθ} \d θ² + g_{ϕϕ} \d ϕ² + 2g_{tϕ} \d t \d ϕ\,,
\end{equation}
null geodesics are solutions to a set of second order ODEs:
\begin{equation}
    \ddot{x}^μ + Γ^μ{}_{νρ} \dot{x}^ν \dot{x}^ρ = 0\,,
\end{equation}
where $Γ^μ{}_{νρ}$ is the torsion-free affine connection, and dots represent derivatives with respect to the affine parameter. These are four second order equations for the four components of the position vector of the photon, which can be integrated given the initial position and velocity of the photon. However, stationarity and axisymmetry imply that energy ($E$) and angular momentum ($L$) are constant along the geodesic. Using these constants of motion, we can integrate the $t$ and $ϕ$ components of the geodesic equation once, to reduce them to two first order equations:
\begin{equation}
    \dot{t} = \frac{E g_{ϕϕ} + L g_{tϕ}}{g_{tϕ}² - g_{tt} g_{ϕϕ}},\quad
    \dot{ϕ} = -\frac{E g_{tϕ} + L g_{tt}}{g_{tϕ}² - g_{tt} g_{ϕϕ}}\,.
\end{equation}
We now have a system of $2$ first order and $2$ second order ODEs, which require $6$ inital conditions to solve. In addition, we have the constraint for a null geodesic, $\dot{x}^μ \dot{x}_μ = 0$, which we use to monitor the accuracy of the integration.

A useful way to parametrize the initial condition is in terms of position and velocity of the photon at the observer's screen. We use the formulation in \cite{Younsi:2016azx}, and place the observer at $r_{\textrm{obs}}=100M$ from the center of the coordinate system, at an inclination of $17^\circ$ and $60^\circ$ respectively. To perform the integration, geodesics are evolved from the observer's screen until they either (i) arrive at the shell (with a tolerance of $r = r_{\textrm{shell}} + 10^{-5}M$), or (ii) exit the camera sphere $r>r_{\textrm{obs}}$.

\subsection{Lensing bands and critical curves}
To understand the image produced on the observer's screen, it is useful to label each photon by the number of times it crosses the equatorial plane.\footnote{For a stationary axisymmetric object, the equatorial plane (which is perpendicular to the spin axis) provides a well defined notion of plane crossing, and in Boyer-Lindquist coordinates corresponds to $θ=π/2$, providing a simple way of tracking photon trajectories. Other notions of $n$ have also been used in the literature, which include counting the number of full turns made by the photon \cite{Gralla:2019xty}, and the number of times the photon crosses a plane perpendicular to the observer's axis \cite{Chael:2021rjo}. For a discussion, see \cite{Cardenas-Avendano:2023obg}.} Photons which cross the equatorial plane exactly once constitute the \emph{direct image} of the object. Each subsequent turn that the photons make around the equatorial plane give a lensed image of this direct image and are referred to as the \emph{lensing bands} denoted by the additional number of times they cross the equatorial plane -- the $n$-th lensing band includes photons that cross the equatorial plane at least $n+1$ times. Each subsequent lensing band becomes exponentially thinner, and the $n→∞$ image coincides  with the \emph{critical curve}, which gives the location of the unstable photon orbit. The complete image contains all the lensing bands superposed on top of one another. By definition, each lensing band includes all the subsequent lensing bands within it, and therefore the critical curve is the brightest curve in the image. The innermost region of the direct image, which corresponds to photons that fall on to the black shell (or hit the horizon of a black hole), therefore remains empty and is referred to as the \emph{inner shadow} \cite{Chael:2021rjo}.
The shape of the critical curve can be conveniently characterized using the \emph{projected diameter} and \emph{centroid} defined by \cite{Gralla:2020yvo}.

\subsection{Ray tracing: results}
With this, we can now study the lensing bands and critical curve for a black shell rotating moderately fast. We will use the rotating black shell metric obtained in \cite{Danielsson:2023onu} to order $a^6$ in Boyer-Lindquist coordinates and perform the computation at $a=0.45$ (corresponding to $a^6 \sim 0.45^6 <1\%$.)

Before presenting the results of the computation, let us first discuss our expectations.
The black shell has two important differences from a Kerr black hole with the same mass and spin: (i) it has a surface which lies outside the outer horizon for a Kerr black hole, and (ii) it has a multipolar structure that differs from Kerr at a percent level. We therefore expect to see two types of differences between the curves of a black shell and a black hole:
\begin{enumerate}[leftmargin=15pt, label=(\roman*)]
    \item The inner boundary of the direct image (equivalently, boundary of the inner shadow) consists of photons that barely escape the black shell (or the horizon of a black hole). A surface that is larger than the horizon should lead to a larger inner shadow. The lensing bands are higher-order lensed images of the direct image -- the inner boundary in particular is a lensed image of the boundary of the inner shadow. We would therefore expect the image of the black shell to have larger inner boundaries for each band, as compared to a black hole.
    
    \item The shape of the critical curve and the outer boundaries of the lensing bands are a function of the spacetime metric outside the object. Since the black shell has a different multipolar structure compared to a Kerr black hole, we expect the shape of the curves to be different. The leading multipolar deviations are small: $\sim 1\%$ for the quadrupole, and $\sim 6\%$ for the octupole and we expect the critical curve to have a small difference in shape compared to Kerr.
\end{enumerate}

\begin{figure*}
    \centering
    \subfigure[Kerr, $θ_{\textrm{obs}}=17^\circ$]{
    \includegraphics[width=0.3\linewidth]{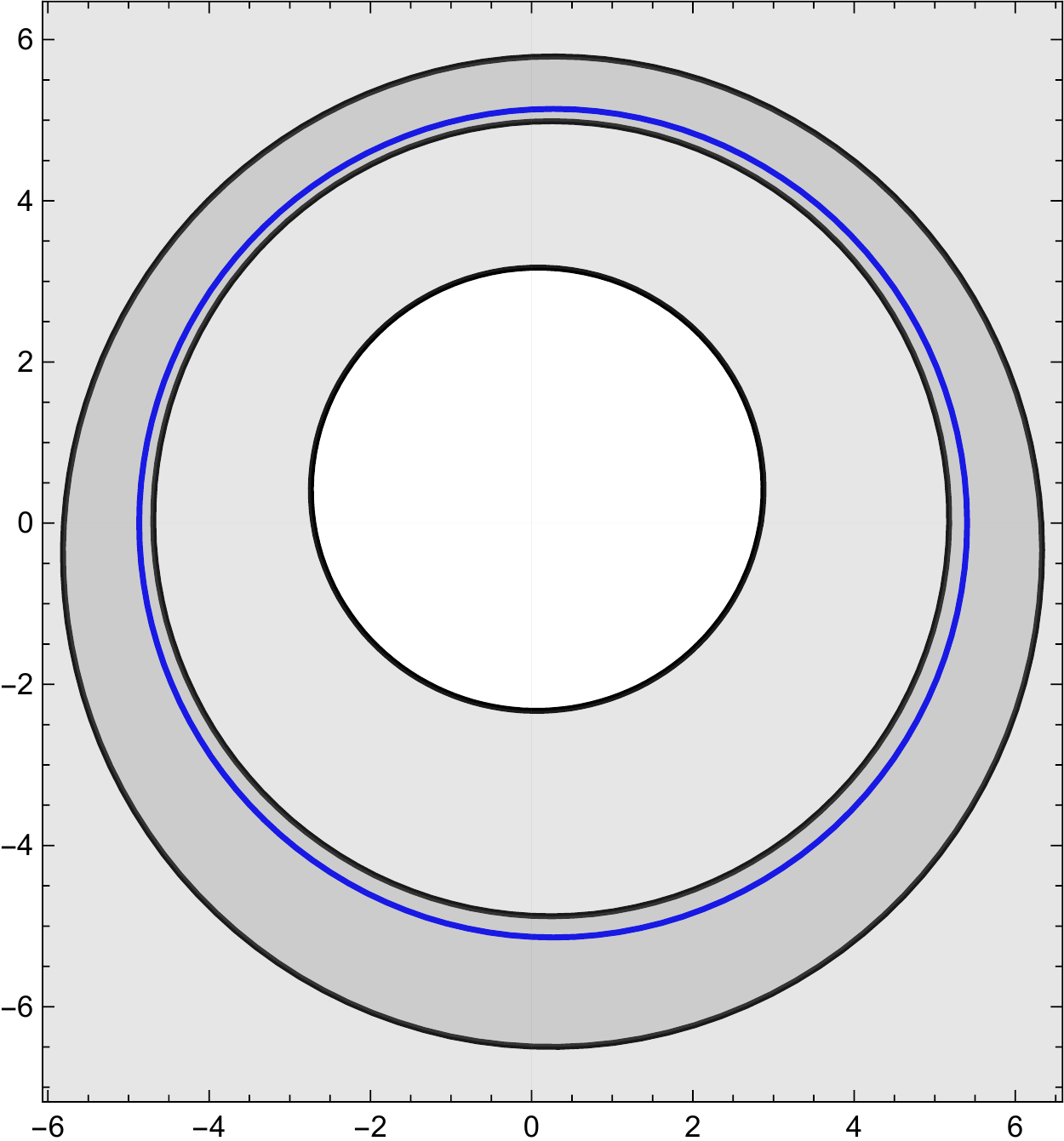}
    \label{fig:kerr17_lensing_bands}
    }
    \hfill
    \subfigure[Black shell, $θ_{\textrm{obs}}=17^\circ$]{
    \includegraphics[width=0.3\linewidth]{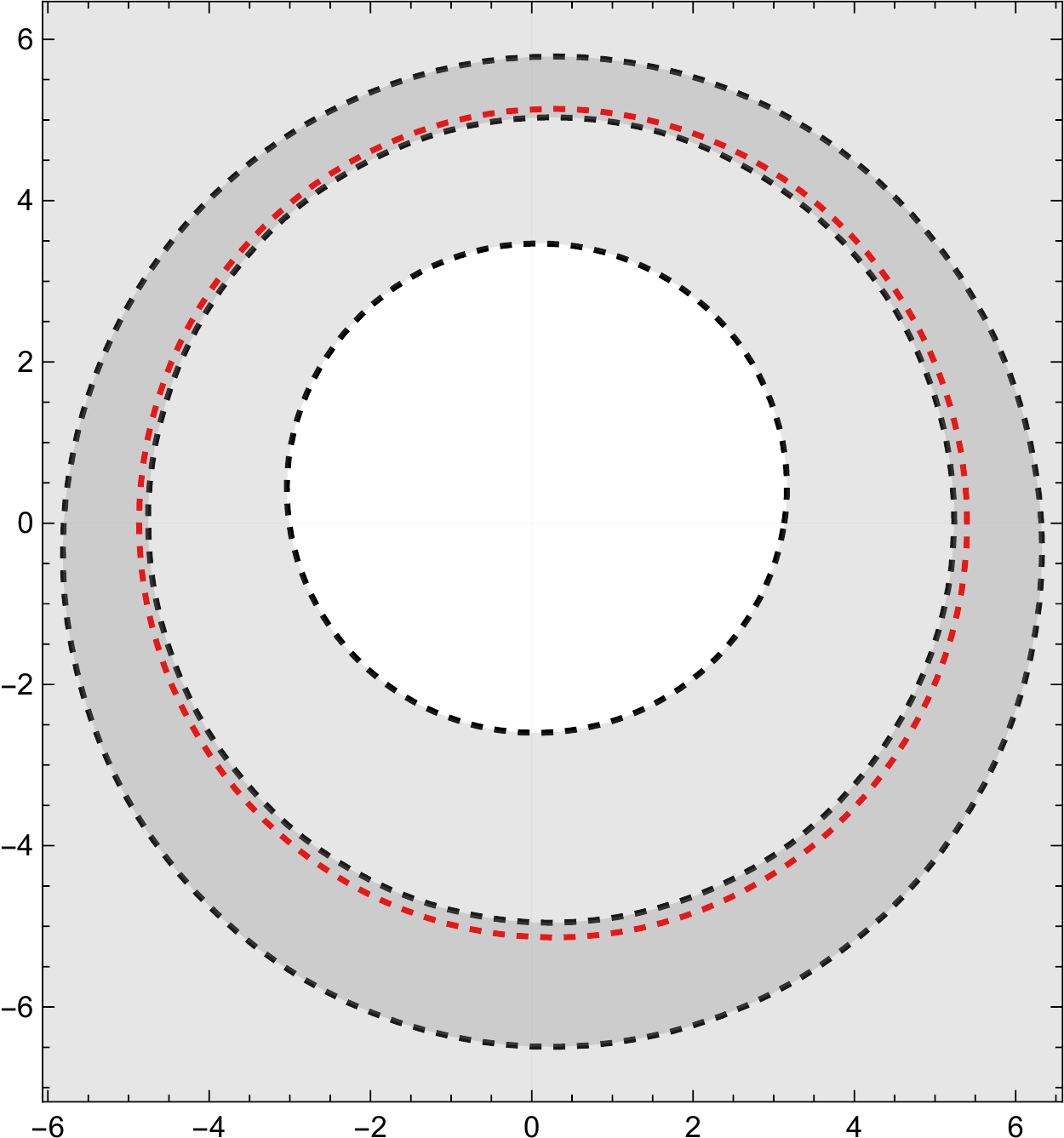}
    \label{fig:BB17_lensing_bands}
    }
    \hfill
    \subfigure[figs. (a) and (b) superimposed]{
    \includegraphics[width=0.3\linewidth]{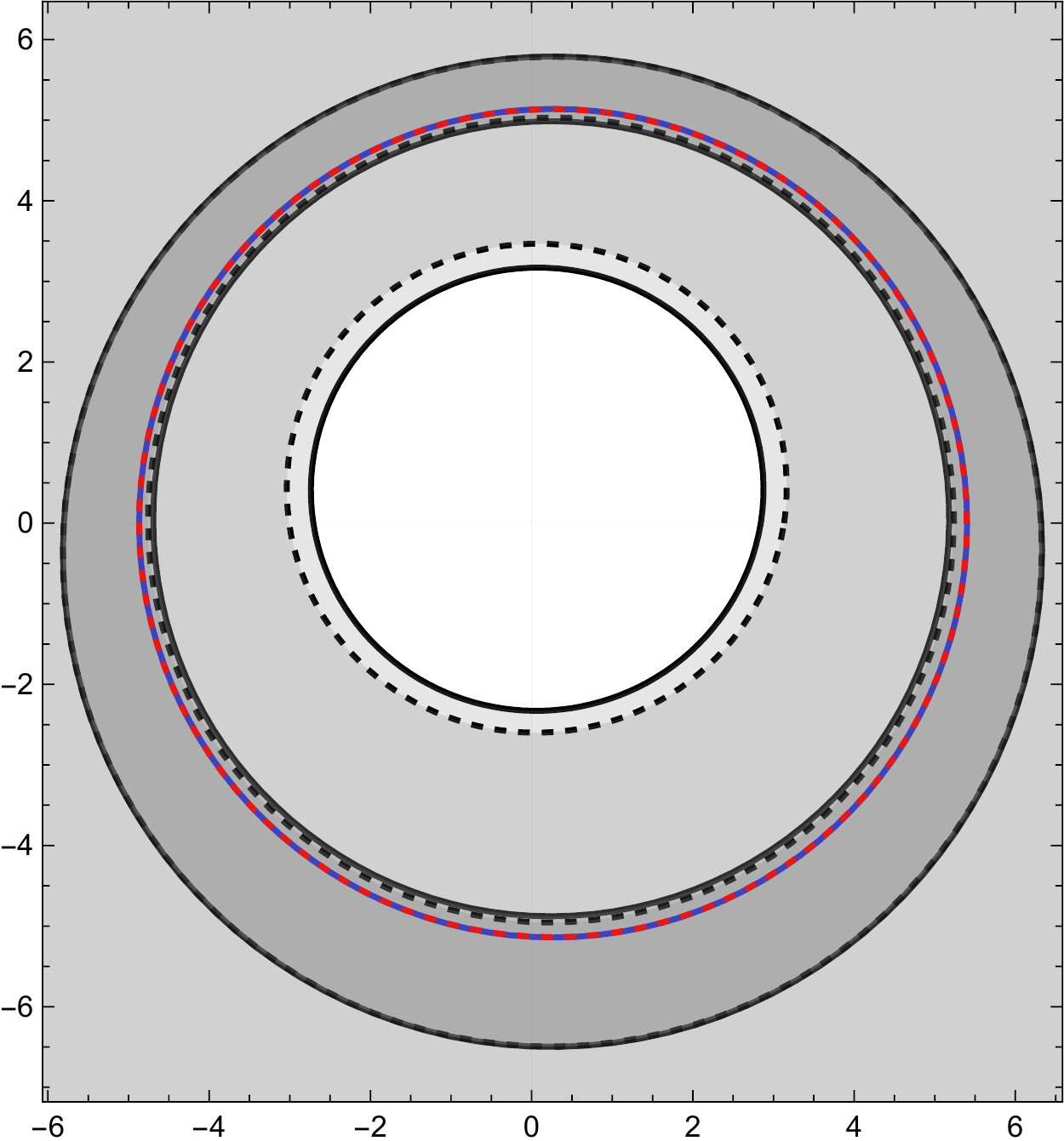}
    \label{fig:difference17_lensing_bands}
    }
    \hfill
    \subfigure[Kerr, $θ_{\textrm{obs}}=60^\circ$]{
    \includegraphics[width=0.3\linewidth]{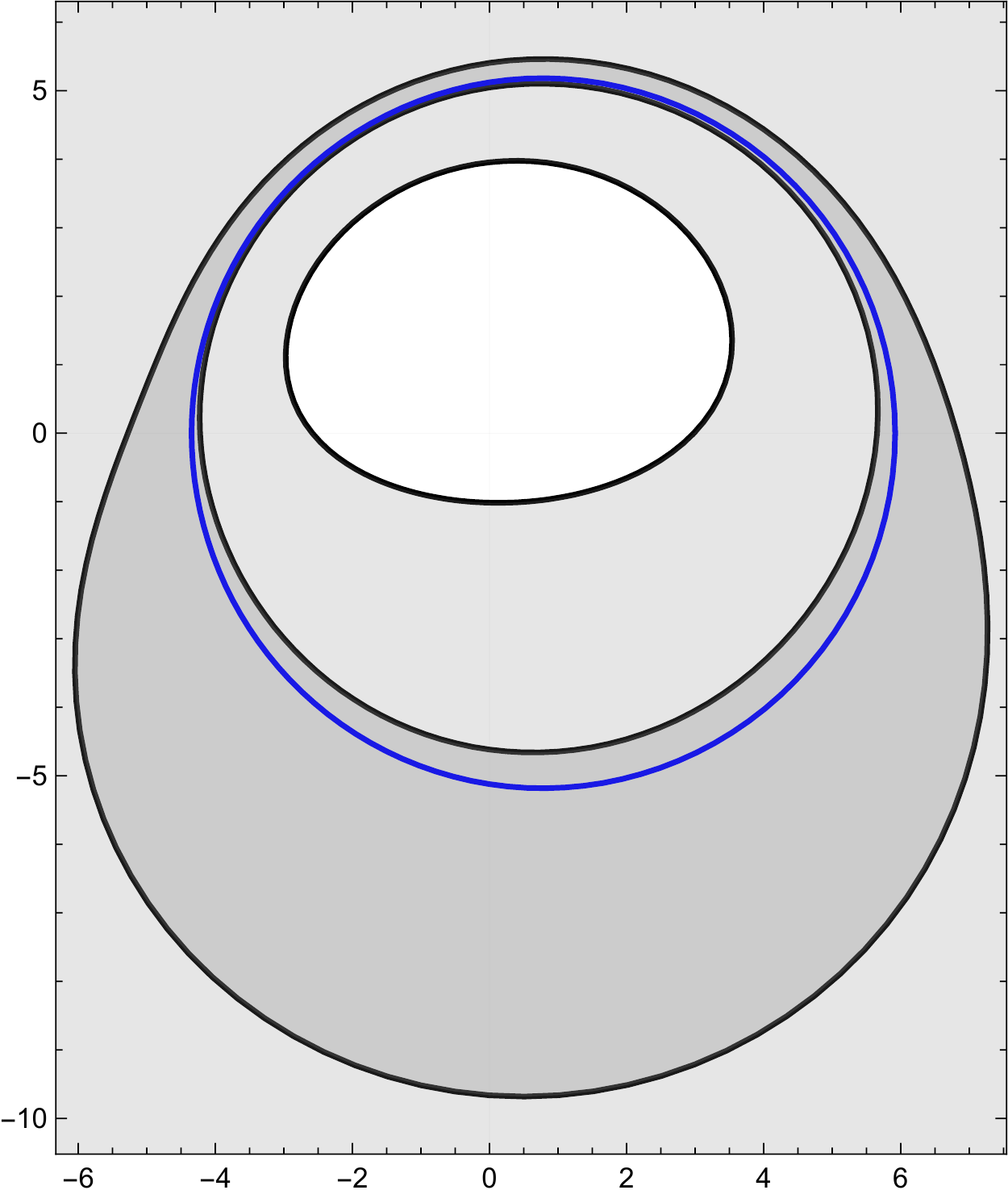}
    \label{fig:kerr60_lensing_bands}
    }
    \hfill
    \subfigure[Black shell, $θ_{\textrm{obs}}=60^\circ$]{
    \includegraphics[width=0.3\linewidth]{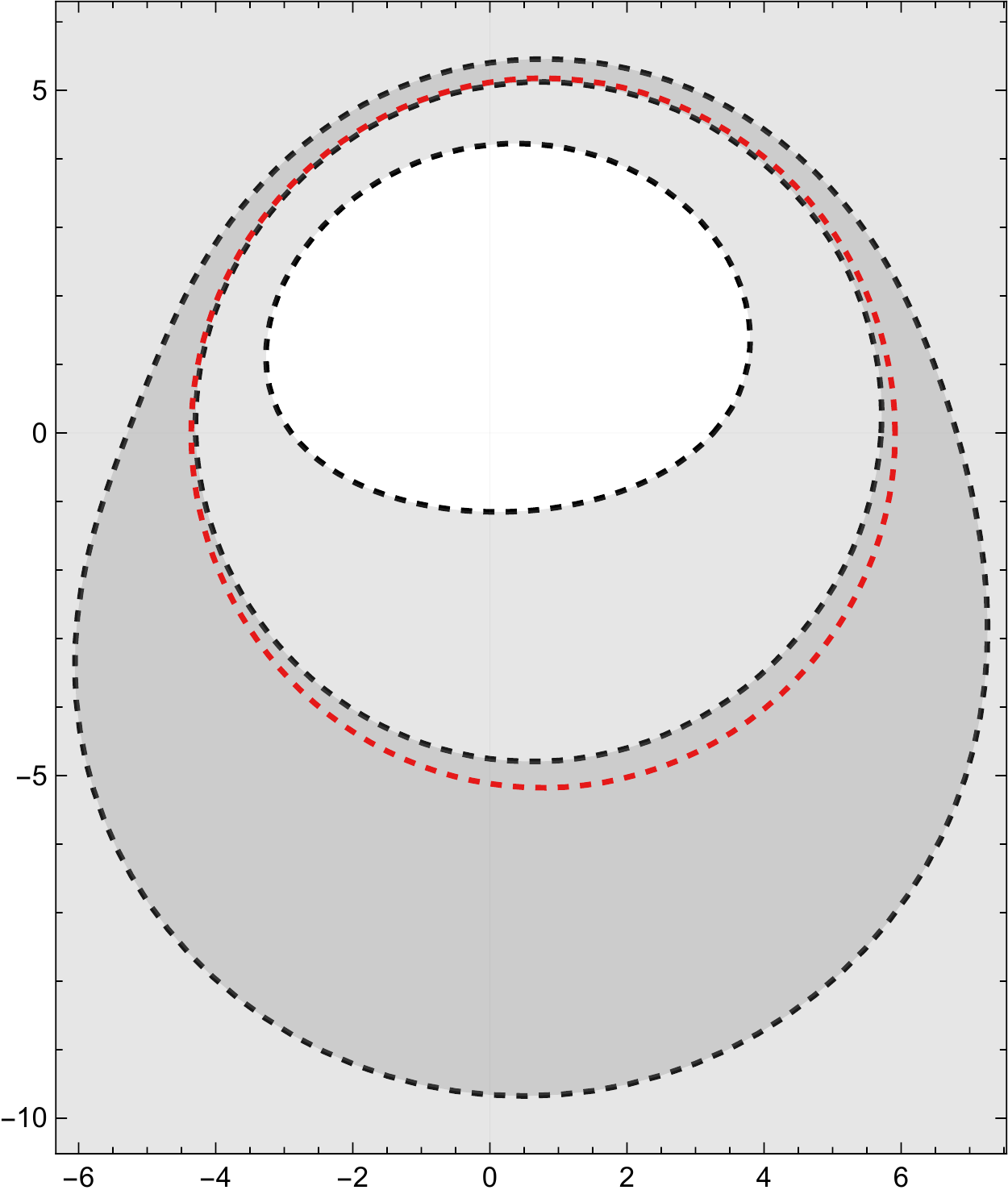}
    \label{fig:BB60_lensing_bands}
    }
    \hfill
    \subfigure[figs. (d) and (e) superimposed]{
    \includegraphics[width=0.3\linewidth]{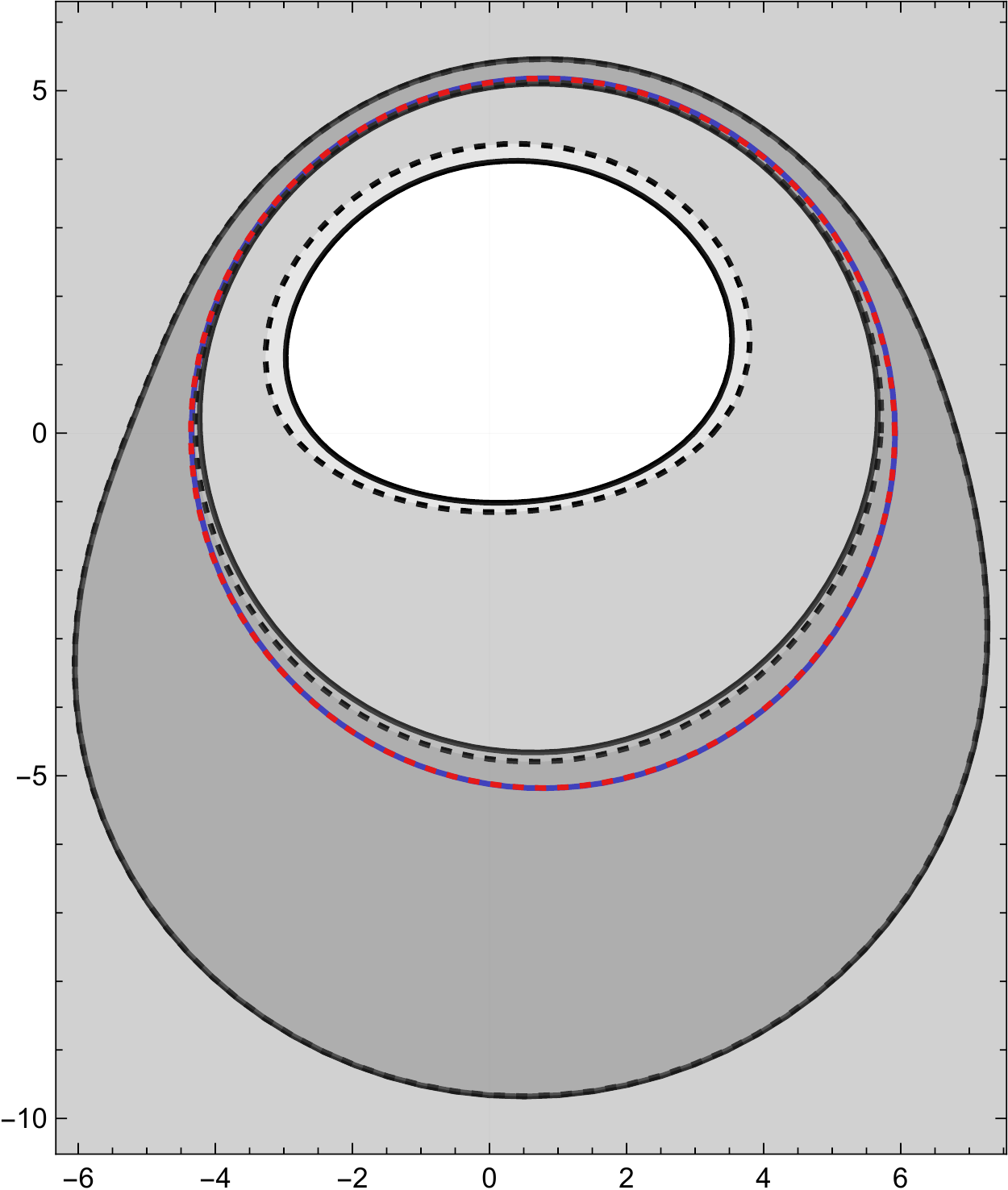}
    \label{fig:difference60_lensing_bands}
    }
    \caption{The direct image ($n=0$) and  the $n=1$ lensing band (light and dark gray regions, respectively) for a Kerr black hole and a black shell as observed from an inclination angle of $17^\circ$ and $60^\circ$ respectively, relative to the spin axis (dimensionless spin of $a=0.45$) are shown in figs. [(a), (b)], and figs. [(d), (e)]. The axes are labeled in units of $M$. Critical curves for the Kerr black hole and the black shell (corresponding to the $n→∞$ lensing band) are shown with solid blue and dotted red lines, respectively. Figs. (a) and (b) are superimposed in fig (c), and figs. (d) and (e) are superimposed in fig. (f). The edges of the lensing bands for the Kerr black hole and the black shell are shown with solid and dotted lines respectively.}
    \label{fig:lensing_bands}
\end{figure*}
\begin{figure*}
    \centering
    \includegraphics[width=\linewidth]{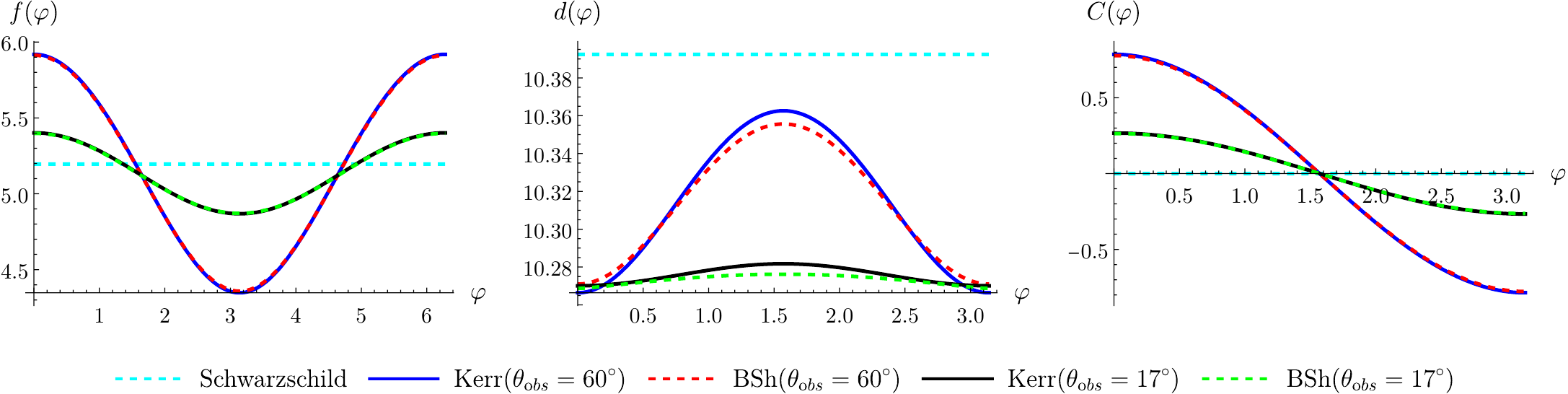}
    \caption{Shape of the critical curve for a Kerr black hole and a black shell (BSh) of the same mass and spin ($M=1, a=0.45$), as seen by observers at two different inclinations with the spin axis, characterized using quantities defined by \cite{Gralla:2020yvo}. The projected position $f(φ)$ as a function of the normal angle $φ$ fully parametrizes the shape of the curve. Even and odd parity components of $f(φ)$ give the projected diameter $d(φ)$ and centroid $C(φ)$ respectively.}
    \label{fig:characteristics}
\end{figure*}

Finally, we can integrate the geodesic equations numerically from the observer's screen with initial conditions given by \cite{Younsi:2016azx}. Since we are only interested in the inner and outer boundaries of the lensing bands, we use a binary search algorithm to find each boundary (\emph{method of bisection}). For the direct image ($n=0$), and the $n=1$ lensing band, we only need to track photons that cross the equatorial plane once and twice respectively. Having obtained those, 
we directly compute the critical curve (corresponding to the $n→∞$ photon ring), also using a binary search (since the higher order lensing bands approach the critical curve exponentially fast).

Before looking at the results, it is useful to look at \cref{fig:curves} to orient ourselves. There, we have produced a simulated 230 GHz synchrotron emission image from an accretion disk made of magnetized plasma around a rotating Kerr black hole with $a=0.45$ using force-free GRMHD and radiative transfer \cite{Gammie:2003rj,Moscibrodzka:2017lcu}.
We have labeled the critical curve, inner shadow and an instance of the projected diameter on the image.

We are now ready to take a look at our results in \cref{fig:lensing_bands,fig:characteristics}. From \cref{fig:lensing_bands}, we see that both of our expectations are realized. In the right most panels, we see that the inner boundaries of the $n=0$ and $n=1$ bands for the black shell are larger than those for a Kerr black hole with the same mass and spin; the difference is big enough to see by eye. The outer boundaries and the critical curve are also not identical, but the difference is much smaller and harder to make out by eye, just as expected.

In \cref{fig:characteristics}, we have computed the projected position, diameter and centroid of the critical curve as a function of the angle. The largest difference is in the shape, given by the projected diameter $d(φ)$, shown in the center panel just as expected. 

While the critical curve for a black shell has a shape distinct from a black hole with identical mass and spin, it is imperative to keep in mind that current uncertainties in mass and spin estimation from astrophysical observations can conflate the shape of the curves, making them indistinguishable~\cite{EventHorizonTelescope:2019ggy,EventHorizonTelescope:2022xqj}.

The size of the inner shadow, and the inner boundary of the $n=1$ lensing band, on the other hand, are the biggest differences between a black shell and a black hole.
Next generation radio-VLBI observatories like the ground based next generation EHT (ngEHT) \cite{Ayzenberg:2023hfw} and the proposed space-ground Black Hole EXplorer (BHEX) \cite{Gralla:2020srx,Johnson:2024ttr} are expected to have a higher resolution (of up to a few μas), and have an improved dynamic range, which could result in the measurement of the $n=1$ lensing band. However, this is far below the resolution required to distinguish a black shell from a black hole in \cref{fig:lensing_bands}. The inner shadow, on the other hand is not a robust observable feature; its observability depends, among other things, on favourable conditions in the accretion flow and optimum observer inclination. 
However, a favorable detection, especially in the context of time-domain observations with next-generation detectors, could provide strong evidence either in favor of or against a black shell.

There is, however, an additional effect that could affect the inner shadow for a black shell owing to how accreting magnetized plasma interacts with the shell. While a black hole horizon provides purely ingoing boundary conditions for the plasma, the surface of a black shell will interact with the plasma in a much more nontrivial way. Although the black shell is expected to be a very efficient absorber similar to a black hole, the microscopics of how the absorption and energy transfer takes place within the black shell could have an effect on the accretion physics. These effects are beyond the scope of the present work and are being investigated separately using GRHMD simulations. We will report on the results in an upcoming work.

\section{Gravitational wave signatures}\label{sec:gw_observables}

Having looked at electromagnetic signatures of black shells, let us now turn our attention to gravitational observables. We will focus on two in particular: (a) Light ring characteristics and quasinormal modes, and (b) the tidal Love number. 

\subsection{Light ring frequency and QNMs}\label{sec:light_ring}
Quasinormal modes (QNMs) are defined as solutions of the spacetime perturbation equations (with complex characteristic frequencies) with boundary conditions that are purely incoming at the horizon (or the surface of a horizonless compact object) and purely outgoing at infinity. A
tantalizing relation between the QNM frequencies in the eikonal limit and unstable null circular orbits for any static, spherically symmetric, asymptotically flat spacetime has been noted~\cite{1972ApJ...172L..95G,Cardoso:2008bp}
(though, for a cautionary note see~\cite{Khanna:2016yow}):
\begin{equation}
    ω_{\textrm{QNM}} = Ω_c l - i\left(n+1/2\right) |λ|,
\end{equation}
where $Ω_c$ is the orbital angular velocity at the null circular geodesic, $λ$ is the principal Lyapunov exponent that characterizes the deviation of nearby photon orbits around the unstable circular orbit, $n$ is the overtone number, and $l$ is the angular momentum quantum number of the perturbation. Beyond spherical symmetry, it is known for Kerr \cite{Ferrari:1984} and Kerr-Newmann \cite{Berti:2005eb} black holes that equatorial null geodesics can account for $l=|m|$ modes. 

Nevertheless, characteristics of unstable null circular orbits in the equatorial plane carry information about the compact object, and the deviation from their Kerr values could serve as a test of how closely they mimic or differ from a Kerr black hole.
\begin{figure*}
    \centering
    \includegraphics[width=\linewidth]{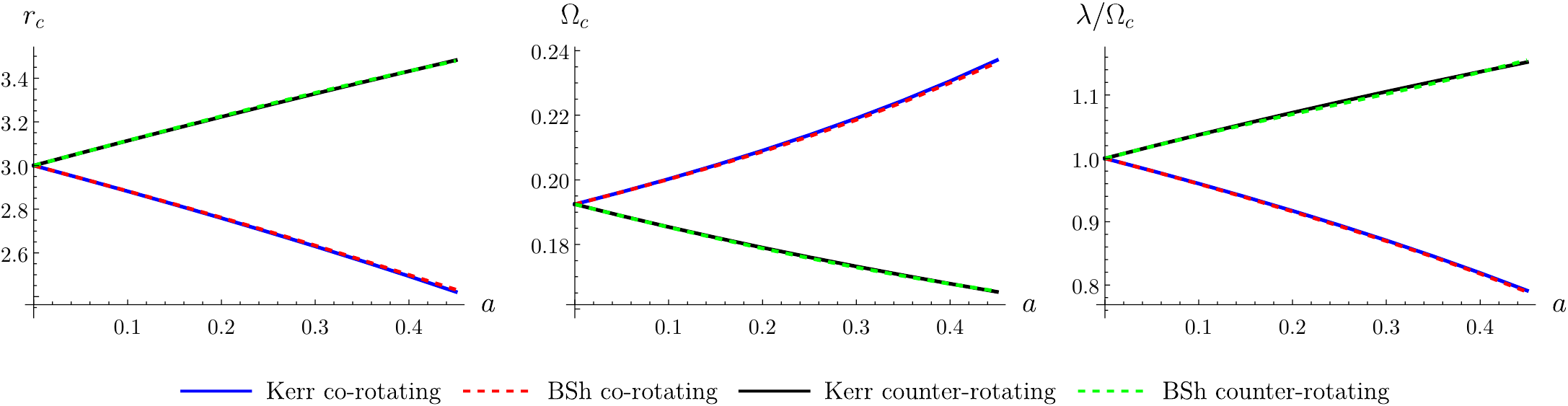}
    \caption{Characteristics of unstable null circular geodesics associated with the black shell (BSh) and a Kerr black hole with the same mass and spin, shown as a function of the dimensionless spin $a$.}
    \label{fig:null_geodesics}
\end{figure*}
Following \cite{Chandrasekhar:1985kt}, let us write down the geodesic equation in the equatorial plane ($θ=π/2, \dot{θ}=0$) and find the radial potential for a stationary axisymmetric metric. The appropriate Lagrangian for the system is
\begin{equation}
    2 \mathcal{L} = g_{tt} \dot{t}² + g_{rr} \dot{r}² + g_{φφ}\dot{φ}² + 2 g_{tφ}\dot{t}\dot{φ}.
\end{equation}
Next, we write down momenta canonically conjugate to the coordinates $p_μ = ∂ \mathcal{L}/∂x^μ$. Stationarity and axisymmetry imply that energy and angular momentum are constants of motion ($p_t = -E, p_φ = L$), from which it follows that $\dot{t}, \dot{φ}$ are given by
\begin{equation}\label{eq:tdot_phidot}
     \dot{t} = \frac{E g_{φφ} + L g_{tφ}}{g_{tφ}²-g_{tt}g_{φφ}}, \quad
     \dot{φ} = -\frac{E g_{tφ} + L g_{tt}}{g_{tφ}²-g_{tt}g_{φφ}}.
\end{equation}
The Hamiltonian corresponding to the system ($2\mathcal{H} = 2\left( p_μ \dot{x}^μ - \mathcal{L} \right) \coloneqq δ \equiv \textrm{constant}$) can be written as
\begin{equation}
    \dot{r}² = \frac{L² g_{tt}+2ELg_{tφ}+E²g_{φφ}}{g_{rr} \left( g_{tφ}²-g_{tt}g_{φφ}\right)} + \frac{δ}{g_{rr}} \coloneqq V_r.
\end{equation}
For null ($δ=0$) circular geodesics to exist at $r=r_c$, $V_r(r=r_c) = V_r^\prime(r=r_c) = 0$.
The orbital angular frequency on these circular geodesics follows from \eqref{eq:tdot_phidot}
\begin{equation}
    Ω_c \coloneqq \frac{\dot{φ}}{\dot{t}}\bigg|_{r=r_c} = -\frac{L g_{tt}+E g_{φφ}}{L g_{tφ}+E g_{φφ}}\bigg|_{r=r_c}.
\end{equation}
The Lyapunov exponent $λ$ is a measure of the instability of the orbit and is given by
\begin{equation}
    λ = \sqrt{\frac{V_r^{\prime\prime}}{2\dot{t}²}}\bigg|_{r=r_c}.
\end{equation}
To each of these quantities, we can attach an associated orbital time scale $T_Ω \coloneqq 2π/Ω_c$, an instability timescale $T_λ \coloneqq 2π/λ$, and define a critical exponent $γ$ following \cite{Pretorius:2007jn}
\begin{equation}
    γ \coloneqq \frac{Ω_c}{2π λ} = \frac{T_λ}{T_Ω},
\end{equation}
or simply a dimensionless instability exponent $λ/Ω_c$.

We can now use the metric of the rotating black shell to order $a^6$ in \cite{Danielsson:2023onu} to compute the orbital frequency and the dimensionless instability exponent for a slowly rotating black shell. It is convenient to introduce the impact parameter $b \coloneqq L/E$. In \cref{fig:null_geodesics} we present the orbital frequency as a function of spin $a$ for both the black shell and a Kerr black hole for the same mass and spin. It is interesting to note that
\begin{equation}
    Ω_c = \frac{1}{b}.
\end{equation}
This special relationship between the angular frequency of null circular geodesics and the impact parameter is known to hold for the Kerr solution \cite{Chandrasekhar:1985kt}. It was discovered in \cite{Cardoso:2008bp} that it holds for a $d$ dimensional Myers-Perry black hole \cite{Myers:1986un} (which is a rotating black hole solution in dimensions $d>4$) as well, where it was further speculated that it holds for any stationary spacetime. That it holds for the black shell provides evidence in support of this speculation. 

The dimensionless instability exponent of a rotating black shell compared to a Kerr black hole of the same mass and spin, as a function of dimensional spin parameter $a$ is shown in \cref{fig:null_geodesics}. Since the rotating black shell metric is known perturbatively to order $a^6$, useful results can only be obtained for spin up to $a \sim 0.45$. The difference from a Kerr black hole is extremely small, at sub-percent to a percent level, for this range of spin.

\subsection{Tidal Love number \& Gravitational Waves}\label{sec:tidal}

Gravitational waves from inspiraling compact objects, prior to the merger
stage, can be obtained through a Post-Newtonian approach.  The system is then
described by the dynamics of two effective ``point particles'' where details
of its structure are captured, to leading order,  for non-spinning objects at the 5th Post-Newtonian
level via a tidal Love number~\cite{Poisson_Will_2014}. 
Such a number helps describe the deformability 
of a self-gravitating object immersed in an external tidal field. In an inspiraling
binary, such a deformation leaves a potentially detectable imprint in the gravitational-wave (GW) signal emitted by the binary in the late stages of its orbital evolution (e.g.~\cite{Flanagan:2007ix}). We here undertake the computation of the tidal Love number for non-spinning black shells. (For simplicity, we do not concern ourselves in this work
with potential modulations introduced by spin-induced quadrupole moments~\cite{Lyu:2023zxv}.)

We follow~\cite{Uchikata:2016qku} and compute the 
(static) tidal Love numbers for non-rotating black shells. 
To this end, we expand the (static) metric as
$g_{ab} = g_{ab}^{(0)} + \delta g_{ab}$, using $(t,r,\theta,\phi)$ coordinates,
\begin{equation*}
\begin{split}
\delta g_{ab} = \mbox{diag}&\left(f H_0^l(r), \frac{H_2^l(r)}{h},r^2 K^l(r),r^2 K^l(r) \sin²θ \right)\times\\
&\times Y_{l0}(\theta,\phi),
\end{split}
\end{equation*}
with $Y_{lm}$ spherical harmonics, and we focus on the spherical symmetric case  background so $m=0$ can be chosen.  The function $f$ is given by
\begin{equation}
f = 1 - \frac{2 \hat M}{r} + \hat \Lambda \frac{r^2}{3},
\end{equation}
with $\hat \Lambda = \Lambda \Theta(R-r)$ and $\hat M = M \Theta(r-R)$ with $\Theta$ the
Heaviside step function and $R$ the shell's radius.
In \cite{Uchikata:2016qku,Pani:2015tga} the case for a gravastar was studied, from which most results can
be derived, as the sign of $\Lambda$ was not crucial. The solution for the fields in each region, at linearized order
imply all functions can be determined from $H_0^l(r)$. This, in turn, obeys
\begin{equation}
\frac{\d^2 H_0^l}{\d r_*^2} - V H_0^l =0 \label{eqnH},
\end{equation}
with the potential
\begin{equation}
V = \frac{9 r^2}{R^4} \left( l (l+1) f + 4 \frac{\hat M^2}{r^2} -
 \frac{2}{9} r^2 \hat \Lambda (9+r^2 \hat \Lambda) \right),
\end{equation}
with the tortoise coordinate defined as $\d r/\d r_* \equiv g(r)= r R^{-2} (6 \hat M - \hat \Lambda r^3 - 3r)$.

Equation (\ref{eqnH}) can be solved independently inside/outside the shell requiring
regularity and asymptotic decay:
\begin{eqnarray}
H_0^{i,l} &=& \alpha_1 \frac{r^l \sqrt{\Lambda}}{3+r^2\Lambda} {}_2F_1\left[\frac{l-1}{2},\frac{l}{2},l+\frac{3}{2},-r^2\Lambda/3\right],\nonumber\\
H_0^{o,l} &=& \alpha_2 P^2_l\left(\frac{r}{M}-1\right) + \beta_2 Q^2_l\left(\frac{r}{M}-1\right),
\end{eqnarray}
where $P^2_l,Q^2_l$ are associated Legendre functions and ${}_2F_1$ the hypergeometric function. The constants $\{\alpha_i,\beta_i\}$ are fixed (up to an overall value) 
by imposing Israel's junction conditions~\cite{1966NCimB..44....1I} (2 equations) 
together with coordinate continuity (1 equation) and the equation of state
describing the shell that modulates the jump at linear order.\footnote{Reference ~\cite{Uchikata:2016qku} works out explicitly for $l=2$ the junction conditions together
with the coordinate condition transformation with respect to time scalings so that the boundary
is at the same coordinate value $r$.}
For concreteness and simplicity we take $\delta p = - \delta \sigma$  (with $p,\sigma$ the pressure and energy density of the shell).
For simplicity we focus on the $l=2$ mode, asymptotically,
\begin{equation}
H_0^2 \rightarrow 3 c_1 \frac{r^2}{M^2} - 6 c_1 \frac{r}{M} + \frac{8}{5} \frac{M^3}{r^3} c_2 + O(r^{-4}),
\end{equation}
where we have introduced the constants $c_i$ which are functions of the
solutions obtained for $\{\alpha_2,\beta_2\}$.  The quadrupole tidal love number is given by \cite{Hinderer:2007mb}
\begin{equation}
\lambda_{\textrm{BSh}} = \frac{8}{45} \frac{c_2}{c_1} M^5.
\end{equation}
The solution depends on $\Lambda$, taking its value to be large (and negative), 
we find to leading order $\lambda_{\textrm{BSh}} \approx 0.27 M^5$. 

In contrast to a Schwarzschild black hole (whose static Love number vanishes), a stationary black shell has a positive Love number. A non-zero Love number will induce a phase shift in the early, low frequency part of a detected gravitational wave signal from roughly comparable mass mergers (the phase shift is suppressed by the third power of the mass ratio), and is potentially a measurable feature of the gravitational waveform (see ~\cite{Flanagan:2007ix}). 
For instance, this has been exploited to scrutinize the neutron star
equation of state in GW170817~\cite{PhysRevLett.121.161101} and
to systematically constrain the Love number for black holes in the LVK 
catalog~\cite{Chia:2023tle}.

To put the non-zero Love number for a black shell into perspective, it is useful to compare it with other compact objects e.g. neutron stars.
They
have $λ \approx 10^{5}$ times larger,
and a proportionately large impact
on observable consequences, e.g. the phase offset
in the gravitational wave.
While a comparable (say in terms of masses) binary neutron star
system merges at a lower frequency (as black shells are significantly more compact) the
overall phase difference ($\delta \Phi)$ due to tidal effects, 
from far separation until ``contact'' is much smaller for
the black shell binary. For instance, in the equal mass case, 
$\delta \Phi_{\textrm{NS}} \approx 10^3 \delta \Phi_{\textrm{BSh}}.$
Naturally, measuring such small phase offsets would require next generation
(3G) detectors.

\section{Summary, Discussion and Outlook}\label{sec:outlook}
Understanding all potential observables related to exotic compact objects
is imperative to scrutinize them  and, in particular, to realize their role to help reveal insights into open problems on the fundamental physics of compact objects. In this work we have studied electromagnetic and gravitational properties of black shells, and discussed the extent to which they are distinguishable from black holes. While we have focused exclusively on black shells as a sub-class of ECOs, another aim of this paper is to provide guidance and lessons for similar effects in the study of other ECOs.

We have analyzed the electromagnetic properties of black shells in the classical limit, and flat spacetime,\footnote{A reasonable approach, since the shell is located at the Buchdahl radius, a point where spacetime curvature effects are weak for supermassive black holes. A more detailed
investigation is certainly doable, as is the case of axisymmetry.} highlighting the lack of reflection for light with wavelengths
commensurate and smaller than the black shell. This is a natural consequence
of such wavelengths being reduced as the wave propagates into a material with high relative permittivity and permeability, and the conductivity lies within a wide interval.\footnote{We have argued in \cref{sec:em_properties} that a large relative permeability and permittivity, $ϵ_r = μ_r \gg 1$, is natural to expect from the stringy nature of black shells. However, we have not yet derived this from first principles, and for the classical analysis in this article, this has served as a well motivated \emph{assumption}. We would like to return to this in the future and attempt to compute/estimate these electromagnetic properties of the shell from string theory.}
To avoid contributing to reflection, the conductivity needs to be small enough, i.e. $\sigma \ll \epsilon \omega$, but large enough to make sure that the electromagnetic skin depth $δ_s$, defined in \eqref{eq:skin_depth}, is much smaller than the thickness of the shell ($δa$), i.e $\delta_s \ll \delta a$, or $\sigma \gg  (2/δa)\sqrt{ϵ/μ}$. For consistency, this requires $ (δa/c_s) \gg (2/ω)$, where $c_s \coloneqq 1/\sqrt{ϵμ}$ is the speed of light within the shell.
This only needs to apply for $\omega \gg c/a$, that is, light with a wavelength smaller than the Schwarzschild radius. This implies
\begin{equation}
    \epsilon_r\, \delta a /a \gg 1 \, ,
\end{equation}
which is the same condition as we derived in \cref{sec:no_hair} around \eqref{eq:no_hair}. In summary, $\sigma$ needs to satisfy
\begin{equation}
   \frac{2}{\delta a}\sqrt{\frac{\epsilon}{\mu}} \ll \sigma \ll \frac{\epsilon c}{a}  \, ,
\end{equation}
Hence, we conclude that all constraints are satisfied if the time it takes for the slowed down light to cross the thin shell (i.e. the black domain) is longer than the light crossing time of the Schwarzschild radius. On the other hand, for processes that are really slow, with time scales orders of magnitude longer than the light crossing time, we could in principle allow for a reduction in the resistance due to an increased skin depth. 
Should precise future estimates of relevant magnetic fields and other plasma characteristics become available, an opportunity to calculate this resistance would emerge, thereby facilitating a meaningful connection with the underlying model.

We have further initiated an effort to connect the black shell model with VLBI observations of supermassive compact objects, and identify key observables that can distinguish them from black holes.  Our first analysis focused on the photon ring, where the differences between a Kerr black hole and corresponding black shell are subtle. Future steps involve investigating astrophysical accretion flow dynamics to achieve a realistic depiction of the image of a black shell. While the accreting plasma falls into the horizon of a black hole, the interaction with a black shell proves significantly more complex. Upon contact with the shell, the incoming fluid transforms into string degrees of freedom, resulting in a novel state of matter with a different equation of state. The nuances of this interaction might reveal characteristics that better distinguish black shells from black holes than the structure of the photon ring.\footnote{
In \cite{EHT_2022_VI,Carballo-Rubio:2023fjj}, attempts have been made to constrain the size and nature of ECOs using EHT data from M87* and Sgr A*, assuming that their surface is in thermal equilibrium with the accreting environment. Since black shells have a very large heat capacity, they are not expected to thermalize and are not bounded by the analysis there.}
Our current efforts are focused on simulating this interaction through a force-free GRMHD simulation of magnetized plasma accreting onto a black shell. We anticipate discussing these findings in forthcoming publications.

We have inferred gravitational wave properties by estimating the QNM frequencies in the eikonal limit (through the photon orbital frequency)
and computing the tidal Love numbers. This complements our previous
study~\cite{Danielsson:2021ykm} which unearthed features related to propagation of scalar waves through the hollow interior of the black shell. Future 
efforts involve extending this to the gravitational wave case, computing the spectrum of black shell QNMs, and contrasting this with that of a black hole. 

A typical characteristic of most ECOs is the presence of a \emph{stable light ring}. In Schwarzschild spacetime, the potential for a radial null geodesic peaks at $r=3M$ (unstable light ring), and declines to $V(r) \rightarrow -\infty$ as $r\rightarrow 0$. In contrast, ECOs typically have non-singular centers, with a potential that rises to $V(r) \rightarrow +\infty$ as $r\rightarrow 0$. This results in a local minimum between $r=0$ and $r=3M$, which allows for a trapping of null geodesics, hence called the \emph{stable light ring}. The presence of a stable light ring is often thought to allow for the possibility of non-linear instabilities, even if the underlying compact object is linearly stable~\cite{Keir_2016}. For a black shell, the stable light ring is situated on the shell. 

While we have not discussed black shell binaries so far, we can make some informed comments on what one would reasonably expect in such a situation. 
Like a binary black hole system, the evolution of a black shell binary would go through roughly three stages: (i) inspiral, (ii) merger,
(iii) ringdown, though arguably, the ringdown phase can be further sub-divided into: \emph{early ringdown}, \emph{ringdown} and \emph{late-ringdown} stages.
\begin{enumerate}[leftmargin=15pt, label=(\roman*)]
    \item We expect the inspiral to be similar to that of a black hole binary, as tidal effects are very small. Furthermore, the high compactness means that up until frequencies $\sim 4/(9 m_T)$, (where $m_T$ is the total mass of the system) the gravitational waves are expected to be very similar. This expectation is supported by non-linear simulations of highly compact boson star merger \cite{Siemonsen:2024snb} as well as of binary black holes
    in higher curvature gravity~\cite{Cayuso:2023xbc}.

    \item The merger is a highly non-linear process, the details of which we cannot yet begin to describe. In \cite{Danielsson:2021ykm}, it was shown how the black shell grows in size under slow accretion of matter. If the accretion is fast enough, or if two black shells collide, we expect tunneling events to take place similar to the nucleation of a black shell when, e.g., a star collapses. The nucleation of a common shell should be a close parallel to the sudden formation of a common apparent horizon when two black holes collide.

    \item During the \emph{early ringdown}, we can imagine that the shell formed from the merger is highly distorted, and will shed its structure on a ``viscous'' timescale. Assuming that the spherically symmetric model analysed in \cite{Danielsson:2021ykm} can inform what happens here, bulk viscosity would play a role here. A back-of-the-envelope estimate gives that the timescale for modes to \emph{smooth out} is of order $ρ_g/(ξ k²)$, where $ρ_g$ is given by the black hole temperature (and commensurate with $m_T$), $ξ$ is the bulk viscosity (which we choose to be $\sim 1$), and $k$ is each wavenumber that can be supported by the black shell (hence they are $ \propto 2π/n $). As a result, high frequencies would smooth out very quickly. Supported sound modes would have frequencies quantized by $1/c_s$, and the \emph{rotating} quadrupole would have a frequency correlated to $Ω_{\textrm{orbital}}$ at the light ring frequency. 

    \item The \emph{ringdown} phase would be marked by quasinormal modes dominating the gravitational waveform. These would be longer wavelength/slower modes, and we would expect their decay rate to be comparable to that of a black hole of comparable mass. The lowest of them are estimated by the light ring frequency as discussed in \cref{sec:light_ring}.

    \item The \emph{late ringdown} would be dominated by the long term behaviour of modes that make it into the shell and leak to the outside. These waves would not be present in a comparable black hole ringdown.
\end{enumerate}
Obtaining an accurate description of all these stages demand significant
new developments, some of which will be presented in due course. Nevertheless, the heuristics given in this work provide valuable guidance moving forward.

We conclude by emphasizing our hope that this work inspires similar efforts involving other exotic compact objects, with the aim to provide as much information as possible to discover or constrain them through observation. Beyond the heuristics presented here, these methods are likely to be relevant and valuable for such an enterprise.

% remove "Acknowledgements" and "References" from table of contents
\addtocontents{toc}{\string\tocdepth@munge}
\begin{acknowledgments}
We are grateful to Alejandro Cárdenas-Avendaño, Daniel Mayerson, 
Paolo Pani, Nils Siemonsen, and George Wong for helpful discussions at various stages of this work. The work of SG was conducted with funding awarded by the Swedish Research Council grant VR 2022-06157.
LL thanks financial support via the Carlo Fidani Rainer
Weiss Chair at Perimeter Institute. LL receives additional financial support from the Natural Sciences and
Engineering Research Council of Canada through a Discovery Grant and CIFAR.
This work was
supported in part by Perimeter Institute for Theoretical
Physics. Research at Perimeter Institute is supported by
the Government of Canada through the Department of
Innovation, Science and Economic Development Canada
and by the Province of Ontario through the Ministry
of Economic Development, Job Creation and Trade. FP acknowledges support from the NSF through the grant PHY-220728.
\end{acknowledgments}

\small
\bibliography{refs.bib}

\end{document}